\def\YES{y}
\def\NO{n}

       \def\InteractiveFigures{\YES}

       \def\NonInteractiveFiguresAvailable{\YES}

\if\InteractiveFigures\YES
  \typein[\FiguresAvailable]{Should the figures be included (y/n)?}
\else
  \def\FiguresAvailable{\NonInteractiveFiguresAvailable}
\fi

\if\FiguresAvailable\NO
  \documentstyle[preprint,aps]{revtex}
\else
  \documentstyle[preprint,aps,psfig]{revtex}
\fi

\tighten

\begin{document}
\preprint{UPR--635--T\hspace{3.5 cm} CERN--TH.7475/94\hspace{3.5 cm}
 hep-th/9411170 \hspace{1 cm}{}}

\draft


\title{Naked singularities in dilatonic domain wall space--times}


\author{Mirjam Cveti{\v{c}}\footnote{Electronic address:
Cvetic@cvetic.hep.upenn.edu}} 
\address{Department of Physics, University of Pennsylvania,
Philadelphia, PA 19104-6396, U. S. A. }
\author{Harald H. Soleng\footnote{Electronic address:
Soleng@surya11.cern.ch}} 
\address{Theory Division, CERN,
CH-1211 Geneva 23, Switzerland}

\date{16 November 1994}

\bibliographystyle{unsrt}

\maketitle


\begin{abstract}

We investigate gravitational effects of extreme,
non-extreme and ultra-extreme domain walls in
the presence of  a dilaton field $\phi$. The dilaton is
a
scalar field without self-interaction that
couples to
the matter potential that is responsible
for the formation of the wall.
Motivated by super\-string
and super\-gravity theories,
we consider both an
exponential dilaton coupling (parametrized with the
coupling constant $\alpha$)\ and
the case where the coupling is self-dual, {\em i.e\/}.\
it has an extremum for a finite value of $\phi$.
For an exponential dilaton coupling ($e^{2\sqrt\alpha\phi}$),
extreme walls (which are static planar configurations with
surface energy  density $\sigma_{\text{ext}}$
saturating the corresponding
Bogomol'nyi bound)\
have a  {\em naked\/}
(planar)\ singularity outside
the wall  for $\alpha>1$,
while for $\alpha\le 1$ the singularity is null.
On the other hand,  non-extreme  walls (bubbles  with two insides  and
$\sigma_{\text{non}}>\sigma_{\text{ext}}$)\
and ultra-extreme  walls
(bubbles of false
vacuum decay
with $\sigma_{\text{ultra}}<\sigma_{\text{ext}}$)\
{\em always\/}
have  naked  singularities.
There are solutions with  self-dual couplings, which
reduce to  singularity-free vacuum domain wall
space--times. However, only non-
and ultra-extreme walls of such a type are
dynamically stable.
\end{abstract}

\thispagestyle{empty}

\pacs{{\mbox{ }}\\
PACS numbers: 04.20.--q\ \ 04.65.+e\ \ 98.80.Cq\\
{\mbox{ }}\\
{\mbox{ }}\\
{\mbox{ }}\\
{\mbox{ }}{\hspace{-1.96 cm}}CERN--TH.7475/94\\
{\mbox{ }}{\hspace{-1.96 cm}}November 1994}

%
%

\renewcommand{\thefootnote}{\arabic{footnote}}
\addtocounter{footnote}{-\value{footnote}}

\section{Introduction}

Domain walls are surfaces interpolating between regions
of space--time with different expectation
values of  some matter field(s).
A domain wall \cite{VilenkinReview} is a vacuum-like
hypersurface where
the positive tension equals the mass density.
In field theory, domain wall
configurations are possible if
the  effective
potential of the  corresponding matter
field(s)\
has  more than one
isolated minimum.
If
each
side of the domain wall
corresponds to  a vacuum, that is, if
on
either side all the
matter
fields have  {\em constant\/}
expectation values, then all the local
properties
of
each side are Lorentz-invariant and each side is a {\em vacuum\/}.
Such domain walls are usually
referred to as  {\em vacuum domain walls\/}.

Vacuum domain walls can be classified according to the
value of their surface energy density $\sigma$, compared to
the energy densities of the vacua outside the wall \cite{CGSLetter,CGS}.
The three types
are: (1)\ extreme walls with
$\sigma=\sigma_{\text{ext}}$ are planar, static walls. In this case
there is a perfect balance between the gravitational mass of the wall
and that of the exterior vacua;
(2)\ non-extreme walls with
$\sigma=\sigma_{\text{non}}>\sigma_{\text{ext}}$
corresponding to non-static bubbles with two centres
and (3)\ ultra-extreme walls with
$\sigma=\sigma_{\text{ultra}}<\sigma_{\text{ext}}$ representing
expanding bubbles
of false vacuum decay.

The extreme domain wall solutions were found \cite{CGR,CG}
in $N=1$ supergravity
theory, representing walls
interpolating between isolated super\-symmetric
vacua of the matter potential.
They correspond to supersymmetric
configurations saturating the corresponding Bogomol'nyi bound.
There are three possible
types \cite{CG} of these
extreme vacuum domain walls. They are classified
according to
the nature of the field path in the
super\-potential or
according to the type of the induced space--times. Type~I
walls interpolate between
super\-symmetric anti-de~Sitter and Minkowski
vacua. Type~II and Type~III
walls interpolate between two super\-symmetric
anti-de~Sitter vacua.  For a Type~II wall
the super\-potential $W(T)$ passes through
zero, and
the conformal factor of the space--time metric
decreases away from the wall on both sides.
For Type~III walls
the super\-potential does not pass through zero, and
the metric
conformal factor is a monotonously increasing function
of $z$, where $z$ maps the spatial
direction perpendicular to the wall.

The global structure of the
induced space--times  of the extreme
domain wall
configurations has been studied in
Refs.~\cite{CDGS,Gibbons,Griffies},
and generalized to and compared with
the non-extreme
ones in
Refs.~\cite{CGSLetter,CGS,Griffies}.
Interestingly, the $(t,z)$ space--time slice
of the Type~I
extreme [non-extreme]\ walls exhibits the {\em same\/}
global space--time structure as the one of the corresponding extreme
[{\em or\/} non-extreme]\ charged black holes;  however,
now the time-like singularity of
the charged  black holes is replaced by another wall.

On the other hand, the existence of dilatons is a generic feature
of  unifying theories, including effective actions of superstring vacua,
certain classes of supergravity theories as well as Kaluza-Klein theories.
`Dilaton' is here used as a generic name for a  scalar
field without self-interactions that couple to the matter
sources, {\em i.e\/}.\ the  potential of scalar
matter  fields as well as the kinetic
energy of gauge fields, and thereby
it modulates the overall strength of such interactions.
In the low-energy effective action of  super\-string theory, the dilaton
plays an essential r{\^o}le
for the ``scale-factor duality'' \cite{Veneziano},
which has been taken as an indication of a ``dual
pre-big-bang'' phase
as a possible alternative to the
initial singularity of the standard cosmological model
\cite{GasperiniVeneziano}.  The
dilaton is believed to play a crucial role in
dynamical supersymmetry breaking as well.

Moreover, in theories with  dilaton field(s),
topological defects in general,
and black holes in particular, have a space--time structure that is
drastically changed compared to the non-dilatonic ones.
Namely, since
the dilaton couples to the matter
sources, {\em e.g\/}.\ the  charge of the
black hole, or it modulates the strength
of
the matter interactions, it in turn
changes the  nature of the space--time.
In the past, charged  dilatonic black holes have been studied
extensively (for a review see Ref.~\cite{Horowitz} and references
therein).

It is therefore of considerable interest to
generalize another type of topological defects, {\em i.e\/}.\
the vacuum domain wall solutions
\cite{VilenkinWeak,BKT,Vilenkin,IpserSikivie,Sato,LCM,Blauetal,CGSLetter,CGS},
by including the dilaton, thus
addressing
the
nature of space--time in the domain wall background with a
varying dilaton field.
Such configurations may be of specific
interest in the study of
domain walls in the early universe as they may
arise in fundamental
theories that include the  dilaton, in particular in
an effective theory from  superstrings.  In addition,  the nature of
ultra-extreme dilatonic  domain walls, which describe
false vacuum decay,  is
of importance in
basic theories  that contain one or more dilaton fields.

The first set of extreme  dilatonic
domain wall solutions~\cite{Cvetic,Cvetic2} was
found in $N=1$
supergravity theory coupled to a linear supermultiplet.
There
the dilaton $\phi$ corresponds
to the scalar component of the  linear
multiplet; $\phi$ couples with an exponential coupling
$e^{2\sqrt\alpha\phi}$ to the
potential of the matter scalar fields, and  with a ``complementary''
exponential coupling
$e^{-2\phi/\sqrt\alpha}$ to the kinetic energy of the gauge fields.
The coupling $\alpha=1$ is the one of the effective tree level action of
the superstring.
The extreme  domain wall configurations for  such actions
correspond to static domain walls
interpolating between supersymmetric minima
of the matter potential.
However,
along with the dependence of the space--time metric on the coordinate
distance from the wall, $z$,
also
the dilaton  field now varies  with $z$.
In
particular, the Type~I walls, which
interpolate between the  supersymmetric
Minkowski space--time, with
matter superpotential  satisfying $W(T)=0$ and
a constant dilaton on the one side, and a
new type of supersymmetric  space--time
where $W(T)\ne 0$ and where the dilaton
varies with  $z$ on the other.
The space--time structure on
the latter side
crucially
depends on the value of the coupling constant $\alpha$.
For $\alpha\le 1$ there is a
planar null singularity, while for $\alpha>1$ the
singularity is {\em naked\/}.  At
the singularity both
the dilaton field and the space--time curvature diverge.

In the $(t,z)$ slice,
the global
space--time structure
of the Type~I extreme dilatonic walls with
coupling $\alpha$  turns out to be
the same \cite{Cvetic2} as the
$(t,r)$ slice of the extreme charged
dilatonic black holes with coupling $1/\alpha$.
The  complementarity  between the
global space--time structure of the extreme dilatonic
domain walls with coupling $\alpha$ and
extreme charged dilatonic black holes
with coupling $1/\alpha$ can be
traced back to  the nature of the coupling
$e^{2\sqrt\alpha\phi}$ of
the dilaton to the matter potential (the source  for the wall)\
and the
complementary coupling $e^{-2\phi/\sqrt\alpha}$ of
the dilaton to the  gauge kinetic energy  (the source of
the charge of the black hole).
The newly found complementarity
($\alpha\leftrightarrow 1/\alpha$)\ between  the
extreme wall and extreme charged
black hole solutions is a generalization of
the one found \cite{CDGS}
 between  extreme
vacuum domain walls ($\alpha=0$)\ and ordinary extreme black holes
($\alpha=\infty$). Interestingly, only for the
$N=1$ supergravity with the
coupling $\alpha=1$,
which corresponds to an effective tree level theory
from super\-strings, are both extreme dilatonic walls {\em and\/}
extreme  charged
dilatonic black holes void of  naked singularities.

In the present paper we
further
investigate
dilatonic
domain walls.  Within $N=1$ supergravity
we  generalize the extreme dilatonic
solutions  to the case  with  an  arbitrary
separable  dilaton K\" ahler potential.
We derive the corresponding
Bogomol'nyi equations for the static, supersymmetric walls,
and in the thin wall approximation also the
energy density of such walls,  which
saturates the corresponding
Bogomol'nyi bound.
Since the form of the separable
K\" ahler potential is kept arbitrary, the analysis applies
also to the self-dual
case, {\em i.e\/}.\ when the K\" ahler potential  has
an extremum for a finite dilaton value.

The major part of this paper involves a study
of the space--time  for  non-extreme  and
ultra-extreme  dilatonic domain walls
in the thin wall
approximation.
The
solutions are non-static bubbles.
These
solutions can be parametrized
by  a parameter $\beta$  that measures
deviation from  extremality;  that is to say, $\beta=0$
represents the corresponding extreme solution.
Within $N=1$
supergravity such
walls would correspond to solutions which
interpolate between two isolated
minima of the matter potential,
where at least one of the isolated minima
breaks supersymmetry.
These walls are  generalizations
of the corresponding  non- and ultra-extreme
vacuum domain walls, but
now, only numerical solutions have been obtained.
For this reason,  the analysis can be done
only for walls for which the boundary
conditions  for the dilaton
field and the metric can be specified uniquely at the
 wall surface.
Nonetheless, such cases include the physically interesting
example of
Type~I walls, which interpolate
between Minkowski space--time with a constant
dilaton  value and
a new type of space--time with varying dilaton,
as well as reflection-symmetric
(non-extreme)\ walls.

Interestingly, the
space--time induced in the dilatonic domain
wall backgrounds can be related to
certain cosmological solutions by
a complex coordinate transformation where
$z$ is replaced with a cosmic time coordinate
and where the potential changes sign.
The non-extremality parameter $\beta$ then plays the r{\^o}le of
a cosmological spatial curvature:
$\beta^2\rightarrow -k$. Thus, the nature of the dilatonic domain wall
space--time
solutions can be related to the corresponding cosmological solutions, and
where it is possible, we draw the analogy.

We concentrate on the case  with
a general exponential  dilaton coupling ($e^{2\sqrt\alpha\phi}$)\
to the  matter potential that creates the wall.
It turns out that---unlike the
extreme case---for any $\alpha$
and any non-zero value of  the non-extremality parameter
$\beta$, there is a {\em naked\/}
singularity on (at least)\
one side of the wall.
We further generalize the
solutions with an arbitrary dilaton coupling function
$f(\phi)$ for the dilaton coupling to the matter potential.
For the
functions
which have an extremum
$\left.df/d\phi\right|
{\mbox{\raisebox{-0.8ex}{{\tiny{$\!\!\phi_0$}}}}}
=0$,
{\em i.e\/}.\ self-dual functions,
one finds
solutions with  singularity-free vacuum domain wall space--times.
However, it turns out that  only  non- and
ultra-extreme walls of this
type are dynamically
stable.
We also comment on the  effects of a dilaton mass,
which in basic theory can be
induced as a non-perturbative effect. Such a
mass (or any other attractive self-interaction)\
does not alter the space--time  sufficiently  to remove
the naked singularity.

The nature of the  space--time for
non- and ultra-extreme dilatonic domain walls,
which  possesses naked singularities,
poses serious
constraints on the phenomenological viability of
theories with dilaton
fields, including a large class of $N=1$ supergravity theories
as well as the  effective
low energy
theory from superstrings.

The paper is organized as follows.
In Section~\ref{Sect:ThinWall} we spell out the
formalism
for the study of  space--time of
dilatonic domain walls in the thin wall
approximation.
Section~\ref{Sect:Supergravity}
contains
the formalism for embedding extreme solutions into
tree level $N=1$ supergravity
theory.
In Section~\ref{Sect:Explicit} we present
the explicit form of the extreme dilatonic domain wall
system
and comment on their
physical properties.
Then, in Section~\ref{Sect:Non/Ultra}
we analyse the non- and ultra-extreme solutions.
Our results are summarized and discussed in the concluding
Section~\ref{Sect:Conclusion}.

\section{Dilatonic domain walls: Thin Wall Formalism}
\label{Sect:ThinWall}

A generic feature of the dilaton is
that it couples coherently to matter
sources, like kinetic energy of the gauge
fields and the potential associated
with the scalar matter fields.
In the case of a domain wall,
due to the dilaton
coupling to the potential of the wall-forming scalar field,
the dilaton will in general vary with the
distance from the wall.
This in turn implies that the gravitational field outside the
wall is determined not only by the direct gravitational effect
of the wall, but also indirectly through the effects
of the varying dilaton, which
in turn  change the
nature of space--time on either side of the wall.

Our main goal is to understand
the global structure of the induced space--times
for such dilatonic domain
wall configurations.
The ones of most interest are the walls
where on one side the dilaton is constant and space--time
is Minkowskian.
The
other side would involve a
new  space--time with a varying dilaton field.
In this paper we shall
primarily
concentrate
on such domain walls, however,
we shall also discuss domain walls,
which are reflection-symmetric.

In this Section
we present the thin wall formalism for the study
of the  induced
space--time outside the wall
region for  any type (extreme, non- and
ultra-extreme)\
of  dilatonic domain walls.
The Lagrangian for the
domain wall system
is specified in
Subsection~\ref{Sub:Lagrange}.
In
Subsection~\ref{Sub:FieldAnsatze}
we specify the Ans\" atze for the metric and
dilaton field and the
resulting
field equations.
The  boundary conditions, in the thin
wall approximation, are given in Subsection~\ref{Sub:Boundary}.
Finally, in Subsection~\ref{Sub:Cosmology}
 we point out a correspondence
between the  dilatonic domain wall systems and
cosmological solutions with a dilaton field.

\subsection{Lagrangian}
\label{Sub:Lagrange}

The starting point for the study of the dilatonic domain walls is the
bosonic part of the
action, $S\equiv \int {\cal L}\sqrt{-g}d^4 x$,
for the space--time metric $g_{\mu\nu}$, the
matter field $\tau$ responsible
for the formation of the wall\footnote{For the sake
of simplicity we introduce only
one scalar field responsible for the formation of the wall. We also
assume that the wall
has no  charge, and thus the gauge fields are turned off
as well.} and the  dilaton field $\phi$,  which
couples to  the matter potential.
In the  Einstein frame the Lagrangian
for the Einstein-dilaton-matter system
is:\footnote{Throughout the paper
we use units such that $\kappa\equiv 8\pi G = c=1$.
Our sign convention for the metric, the Riemann tensor,
and the Einstein tensor,
is
of the type $-$~$+$~$+$
as classified by Misner, Thorne and Wheeler \cite{MTW}.
}
\begin{equation}
{\cal{L}} =
-\frac{1}{2}R + \partial_{\mu}\tau\,
\partial^{\mu}\tau+ \partial_{\mu}\phi\,
\partial^{\mu}\phi
-V(\phi,\tau).
\label{Lagrangian}
\end{equation}
The potential $V(\phi,\tau)$ is of the form:
\begin{equation}
V(\phi,\tau)=f(\phi) V_{0}(\tau)
 + \widehat V(\phi).
\label{potential}
\end{equation}
The  dilaton modulates
the matter potential $V_0(\tau)$ with the
function  $f(\phi)$ which  we shall denote ``the dilaton coupling''.
For the sake of generality
we  have also added a
self-interaction term  $\widehat V(\phi)$  to the potential
(\ref{potential}). This term is not there in the original
theory.  It is, however, believed to be  generated
after dynamical  (super)symmetry breaking,
and it is responsible for giving
the
dilaton a mass.

In general, the potential (\ref{potential})\
need not
have a supersymmetric embedding.
In this case the walls  are not static, in general.
Due to the complexity of the field equations,
we address the  space--time properties
in the thin wall approximation, only.
Namely, in this case we treat the wall as infinitely thin. It is
located at, say,
$z=0$ (in the rest frame of the wall).  Inside the wall,
the matter field $\tau$, which makes up the wall,  yields the
stress-energy of the wall
with the following $\delta$-function contribution:
\begin{equation}
T^{\mu}_{\;\;\nu}=\delta(z)\,\sigma\,{\mbox{diag}}(1,1,1,0)
\label{stressen}
\end{equation}
where $\sigma$ is the  energy density per unit area of the wall.
Outside the wall the matter field $\tau$ has
{\em no kinetic energy\/},
and thus its
contribution to the energy-momentum tensor on the two sides
is given by the
constant potentials
$V_{0}(\tau_1)$ and
$V_0(\tau_2)$, respectively.
Across the thin wall region, both the metric and the dilaton are
continuous functions; however, their derivatives are discontinuous.
Since the dilaton couples to the matter potential $V_0$,
on each side of the wall,
$V_{0}$ becomes a factor in
an
{\em effective potential\/}
for the dilaton:
\begin{equation}
V(\phi)=f(\phi) V_{0}(\tau_{i}) +
 \widehat V(\phi)
\label{dilpot}
\end{equation}
where $\tau_{i}$ is given by $\tau_{1}$ and $\tau_{2}$
on the two sides of the wall.
The Lagrangian for the dilaton-metric system outside the wall is of
the form:
\begin{equation}
{\cal{L}} =  -\frac{1}{2}R  +\partial_{\mu}\phi\,
\partial^{\mu}\phi
-V(\phi).
\label{Lagrangianp}
\end{equation}
 In the following Subsection we discuss the field
Ans\" atze, the field equations, and the boundary conditions.

\subsection{Symmetries and field equations}
\label{Sub:FieldAnsatze}
The domain wall configurations are most conveniently
 described in the co-moving frame of the wall system.
With the requirement of homogeneity, isotropy, boost invariance,
and geodesic completeness
of the space--time intrinsic to the wall, and the constraint that the
same symmetries hold in the hypersurfaces parallel to the wall, the
line-element is given as \cite{CGS}:
\begin{mathletters}
\begin{equation}
ds^2=e^{2a(z)}\left( dt^2-dz^2 -\beta^{-2}\cosh^2\!\beta t\,\,
d\Omega_{2}^{2}
\right),
\label{metrica}
\end{equation}
where
\begin{equation}
\beta^{-2}d\Omega_{2}^2\equiv \left\{
   \begin{array}{ll}
      \beta^{-2}d\theta^2 +\beta^{-2}\sin^2\!\theta\,d\varphi^2\ \ &
          {\mbox{if }}\beta\neq 0 \\
      dx^2+dy^2 & {\mbox{if }}\beta = 0.
   \end{array}
\right.
\label{metricb}
\end{equation}
\label{metricansatz}
\end{mathletters}\\[-1.0em]
The $z$-coordinate  maps the direction normal to the domain wall.
For later
convenience the conformal factor in Eq.~(\ref{metrica})\
is chosen to be of the
exponential type  $e^{2\,a(z)}$.
Note that
$\beta$ parametrizes  a deviation of the domain wall
configuration from the corresponding planar,
extreme wall (with $\beta=0$).
In accordance with the symmetries of the
metric, the dilaton field $\phi$ is a
function of $z$, only.
In other words,  the dilaton
field is
``tied''
to the wall system and
can thus  vary only  in a spatial direction
perpendicular to the wall.

The Einstein tensor
metric Ansatz (\ref{metrica})\ is
\begin{equation}
\left. \begin{array}{lcl}
G^{z}_{\;\;z} &=&
 3\,\left( \beta^2 - {{ a'}^2} \right)  e^{-2\,a}\\
G^{i}_{\;\,i}&=&
  \left( \beta^2 - {{ a'}^2} - 2\, a''\right) e^{-2\,a}
       \end{array}
\right.
\label{oGmunu}
\end{equation}
where $a'\equiv da/dz$ and  the index $i$ stands for
a coordinate $x^{i}\in\{t,x,y\}$.

The corresponding energy-momentum
tensor on either side of  the wall is of the form
\begin{equation}
\left.   \begin{array}{lcl}
T^{z}_{\;\; z} &=&
  V(\phi)
  -
 {{{{{\phi'}}^2}} {{e^{-2\,a}}}}\\
T^{i}_{\;\, i} &=&
  V(\phi)
+
   {{{{{\phi'}}^2}} {{e^{-2\,a}}}},
      \end{array}
\right.
\label{oTmunu}
\end{equation}
where $V(\phi)$ is the effective `dilaton potential'
on either side of the wall  as defined
in Eq.~(\ref{dilpot}) and $\phi'\equiv d\phi/dz$. Einstein's field equations
and
the second Bianchi identity then lead
to the following set of field equations
\begin{mathletters}
\begin{eqnarray}
   & &
e^{2 a}\, V(\phi)
  +  2\,{\phi'}^2 + 3\, a'' = 0
     \label{fieldeqs1}\\
  & &
-e^{2\, a}\,
{{\partial V(\phi)}\over{\partial\phi}}
+
   4\, a'\,\phi' + 2\,\phi''=0
     \label{fieldeqs2}\\
  & &3\,\beta^2
 -
e^{2a}\, V(\phi)
- 3\,{ a'}^2 + {\phi'}^2= 0 .
     \label{fieldeqs3}
\end{eqnarray}
\label{fieldeqs}
\end{mathletters}\\[-1em]
Eq.~(\ref{fieldeqs2}), which is the energy-momentum conservation
law, is identical to the field equation obtained by varying the
action with respect to the dilaton. Hence, there are only two
independent field equations.  Eq.~(\ref{fieldeqs3})\
can be used to determine the
boundary conditions.

\subsection{Boundary conditions}
\label{Sub:Boundary}
For the solutions on each side of the wall,
Israel's matching conditions \cite{Israel}
relate the  energy density
(\ref{stressen})\
of the wall
to
the discontinuity of the
first-order derivatives
of the metric in the direction transverse to the singular
surface (see Ref.~\cite{CGS} for
details\footnote{In Ref.~\cite{CGS} the
metric coefficient $A(z)$ is related to $a(z)$ by $A(z)=
e^{2a(z)}$.}):
\begin{equation}
\sigma = \left.2 \zeta_{1}  a'\right|
{\mbox{\raisebox{-1.2ex}{{\tiny{$\!0^{-}$}}}}}
- \left.2 \zeta_{2}  a'\right|
{\mbox{\raisebox{-1.2ex}{{\tiny{$\!0^{+}$}}}}}
\label{metricbc}\end{equation}
where $a'=da/dz$.
Here,  $\zeta$ is a sign factor which is
$+1$ if $a'>0$ and $-1$ if $a'<0$ on the corresponding
side of the wall.
Without loss of generality,  we have  normalized the metric
coefficient at the wall to be $a(0)=0$.

Similarly, integrating the equation of motion for
the scalar field across the wall region and employing  the form
(\ref{stressen})\ of the
stress-energy associated with the wall region, one
finds that  the discontinuity in the first-order
derivative of the dilaton at the wall surface is given by:
\begin{equation}
\left. \phi'\right|
{\mbox{\raisebox{-1.2ex}{{\tiny{$\!0^{+}$}}}}}
-
\left. \phi'\right|
{\mbox{\raisebox{-1.2ex}{{\tiny{$\!0^{-}$}}}}}
=
\left.\int^{0^{+}}_{0^{-}} \!\!\!
e^{2 a} \,V'\, dz  =\frac{\sigma}{2}\frac{\partial\ln[f(\phi)]}
{\partial\phi}\right|
\!{\mbox{\raisebox{-2.2ex}{{\tiny{$\phi_0$}}}}}.
\label{dilbc}
\end{equation}
Here, again, we have used the fact
that the dilaton field is continuous across the
wall,  with the value $\phi(0)=\phi_0$.\footnote{In
the thin wall
approximation  the self-interaction potential
$\widehat V(\phi)$
of Eq.~(\ref{potential})\
does not contribute to the
energy-momentum~(\ref{stressen})\ of  the wall.}

In the case when
the solution for the metric coefficient and the dilaton is
known on one side,
one can use Israel's matching condition (\ref{metricbc})\
and the dilaton matching
condition  (\ref{dilbc})\  to find the
boundary condition for the metric
and the dilaton on the other side of
the wall, and thus the form of the solution on the other side of the wall.
In the case of Type~I walls, the solution on
the side of the wall with
$V_0(\tau_2)=0$
corresponds to the Minkowski space--time with a {\em constant\/}
dilaton field. Then,
the matching conditions  (\ref{metricbc})\ and (\ref{dilbc})\
can be used to determine the solution on
the other side of the wall.

In the case of reflection-symmetric walls, the matching conditions
(\ref{metricbc})\ and (\ref{dilbc})\
fix the boundary conditions on both sides of the wall.
This
enables one to find the reflection-symmetric solutions.

As we shall see in the subsequent sections,   analytic
solutions  outside the wall region have been obtained
only in the case of extreme
(supersymmetric)\
walls. For the non-extreme walls we have obtained
only numerical solutions, in general.
Therefore, the above boundary conditions
for the Type~I and the reflection symmetric walls have been used to
determine the boundary conditions for our numerical integrations.

\subsection{Relationship between the dilatonic
domain wall system and a cosmological model}
\label{Sub:Cosmology}

In this subsection
we show that the  Einstein-dilaton system
outside the  wall is
equivalent to that of an Einstein-dilaton
Friedmann-Lema{\^{\i}}tre-Robertson-Walker (FLRW)\ cosmo\-logy.
Formally, one can flip the wall into a space-like hyperspace
by
a complex coordinate transformation:
\begin{equation}
z\rightarrow \eta,\ \ \ \cosh\beta t\rightarrow i \beta r.
\label{tocosmologyform}
\end{equation}
If one regards the new coordinates
as real, then
$\eta$ becomes the conformal time and $r$ a spatial coordinate in a
metric with opposite signature: $(-,+,+,+)$. Changing the
sign of the metric implies a change of sign of the curvature
scalar,
$R$, and of all kinetic
energy terms in the Lagrangian (\ref{Lagrangianp}).
Because the overall sign of the total Lagrangian is arbitrary,
one can change back
the sign of the metric,
if one at the same time changes the sign of
the potential $V(\phi)$.  Thus, the complex
coordinate transformation (\ref{tocosmologyform})\
{\em maps the domain wall system onto  a cosmological model
having a potential with the opposite sign. }
As a result,
the line-element takes the form
\begin{equation}
ds^2=  e^{2 a(\eta)}\left[ d\eta^2 - \frac{dr^2}{1+\beta^2 r^2} -
r^2 d\Omega_{2}^{2}\right].
\label{FLRWmetric}
\end{equation}
This is a FLRW line-element where
the spatial curvature is $k=-\beta^2$.

The equivalence of the Einstein-dilaton system outside the wall  with the
dilaton-FLRW  cosmology  (by using the
coordinate transformations
(\ref{tocosmologyform}),
as well as identifying
$V(\phi)\rightarrow -V_{\text{c}}(\phi)$ and
$\beta^2\rightarrow -k$) proves useful, because
it allows us to
carry over results from the corresponding  cosmological studies.
In the cosmological  picture the domain wall
is a space-like hyper-space. It could be
interpreted as representing  a
phase transition taking place
simultaneously throughout the whole universe.
In our case, the boundary conditions at this hyper-space are fixed
by the
boundary conditions
of the domain wall.

In addition,  we have found it useful to
compare the Einstein-dilaton system outside the wall with the
evolution  of corresponding well-known
perfect fluid
cosmologies and to compute the corresponding effective equation of state
for the dilaton.
In terms of a perfect fluid description,
the energy-momentum tensor  is of the form:
\begin{equation}
\left. \begin{array}{lcl}
T^{\eta}_{\;\; \eta} &=&
  V_{\text{c}}(\phi) +
   {{{{{\dot \phi}}^2}} {{{e}^{-2\,a}}}}\\
T^{i}_{\;\, i} &=&
 V_{\text{c}}(\phi)-
  {{{{{\dot \phi}}^2}} {{{e}^{-2\,a}}}},
      \end{array}
\right.
\label{Tmunu}
\end{equation}
where $\eta$ is the conformal time and
$\dot \phi=d\phi/d\eta$.
Here the index $i$
refers to the three
spatial coordinates.
Note that in expression (\ref{Tmunu})\ the sign of
$V_{\text{c}}(\phi)$ is
reversed with respect to the potential of Eq.~(\ref{oTmunu}).
The expressions (\ref{Tmunu})\
correspond to an energy
density
\begin{mathletters}
\begin{equation}
\rho\equiv T^\eta_{\;\;\eta}=
\dot{\phi}^2 e^{-2 a}
+
V_{\text{c}}(\phi)
\label{rho}
\end{equation}
and a pressure
\begin{equation}
p\equiv -T^i_{\;i} =
\dot{\phi}^2 e^{-2 a}
-V_{\text{c}}(\phi)
\label{pressure}
\end{equation}
\end{mathletters}
\noindent
of a perfect fluid with a four-velocity
$u^{\mu}=e^{-a}\delta^{\mu}_{\;\;\eta}$.

It is conventional to parametrize the equation
of state of a perfect fluid
by the $\gamma$-parameter: $p = (\gamma-1)\rho$.
The following values of $\gamma$ are singled out:
$\gamma=0$ corresponds to the equation of state of a
cosmological constant;
$\gamma=2/3$ is the equation of state of a cloud of strings;
$\gamma=1$ represents dust (non-relativistic cloud of particles);
$\gamma=4/3$ is radiation (ultra-relativistic matter);
and
$\gamma=2$ corresponds to a Zel'dovich fluid (maximally stiff matter).
All physical equations of state are confined to the range $0\le\gamma\le 2$.
This  is
also the range covered by a minimally coupled scalar field $\phi$:
\begin{equation}
\gamma =  \frac{2 \dot\phi^2 e^{-2a}}{\dot\phi^2 e^{-2 a}
+V_{\text{c}}(\phi)}.
\end{equation}
It has $\gamma=2$
if the kinetic energy dominates, and $\gamma =0$ if the potential
energy dominates.
Matter satisfying an equation of state with $\gamma <1$
has negative pressure.  If $\gamma <2/3$, then the
repulsive gravitational effect of the negative pressure is
greater than the attractive gravitational effect
of the energy density. Matter obeying such
an equation of state is therefore
a source of repulsive gravity. An effective equation of state of this
kind is a necessary ingredient in
inflationary universe models.


\section{Supersymmetric embedding}
\label{Sect:Supergravity}

Solutions of the theory specified by the Lagrangian (\ref{Lagrangian})\
can be embedded in a super\-gravity theory.
In this section we spell out the formalism  for an embedding of dilatonic
domain walls into the corresponding tree level $N=1$ supergravity theory.
Extreme domain walls turn out  to be  static, planar configurations
interpolating between {\em supersymmetric\/} minima of
the corresponding supergravity potential. Such configurations satisfy the
corresponding Bogomol'nyi equations for {\it  any thickness} of the wall and
the
energy density  of the wall, which can be precisely defined only in the thin
wall
approximation,  saturates the corresponding Bogomol'nyi bound.
In the thin wall approximation  the Einstein-dilaton system
outside the wall is described  by the
formalism spelled out in the previous section
with the non-extremality parameter $\beta
=0$.

The results presented in this section is
a generalization of the previous work on supergravity
walls\cite{CGR} without the dilaton, and
dilatonic supergravity  walls
\cite{Cvetic,Cvetic2} with a special
form of the  function $f(\phi)=e^{2\sqrt\alpha
\phi}$ in the dilaton effective potential (\ref{dilpot}).
The latter ones arise in  $N=1$ supergravity
with a  general coupling of the
linear supermultiplet.

In Subsection~\ref{Sub:SupergravityLagrangian}
the supersymmetric embedding of extreme dilatonic domain walls is
realized  within $N=1$
supergravity theory with (gauge neutral)\ chiral superfields whose
K\" ahler and super-potential are constrained.  Namely,
the Lagrangian in such a $N=1$ supergravity
contains (gauge neutral)\
matter chiral-superfield $\cal T$,
whose scalar component
$T$ is  responsible for a formation of the wall. In addition,
there is a chiral superfield $\cal S$,
which has no superpotential and whose K\" ahler potential decouples
from the one of $\cal T$.
In turn, the scalar component $S$ of the chiral
superfield $\cal S$
acts as the dilaton
field, which couples to the matter potential.
We derive the Bogomol'nyi bound and
Bogomol'nyi equations (Killing spinor equations)\
for the extreme
dilatonic  wall solutions  in
Subsections~\ref{Sub:BogomolnyiBound}
and
\ref{Sub:KillingSpinorEqs}, respectively.
The latter subsection also contains a classification
of the extreme dilatonic domain walls.

\subsection{$N=1$ Supergravity Lagrangian}
\label{Sub:SupergravityLagrangian}

In  supergravity theories the bosonic Lagrangian
(\ref{Lagrangian})\
arises in  $N=1$ supergravity when the  gauge-singlet matter chiral
superfield ${\cal{T}}$
has a non-zero K\" ahler potential (real
function of chiral superfields)\ $K_{\text{matt}}({\cal T},{\cal T}^*)$
as well as a non-zero superpotential (holomorphic
function of the fields)\
$W_{\text{matt}}({\cal T})$.\footnote{For the sake of simplicity
we assume that the wall is formed by  a scalar component of one chiral
superfield only.}   In addition, there is
a chiral
superfield  $\cal S$ with the
K\" ahler potential  $K_{\text{dil}}({\cal S},{\cal S}^*)$,
which {\em does not\/}
couple to the matter K\" ahler potential, and with
{\em no\/}
superpotential ($W_{\text{dil}}({\cal S})=0$).
The bosonic part of the Lagrangian
without the gauge fields is then fully specified by:
\begin{equation}
\left. \begin{array}{c}
K=K_{\text{dil}}({\cal S}, {\cal S}^*)+ K_{\text{matt}}({\cal T},{\cal T}^*),\\
W=W_{\text{matt}}({\cal T}),
       \end{array}
\right.
\label{kalpot}
\end{equation}
and
it is given by
\begin{equation}
{\cal{L}} =  -\frac{1}{2}R +K_{TT^*} \partial_{\mu}T\,
\partial^{\mu}T^*+ K_{SS^*}\partial_{\mu}S\,
\partial^{\mu}S^*
-V
\label{susylag}
\end{equation}
where
the potential is
\begin{equation}
V= e^{K}\left[|D_TW|^2K^{T\,T^*}
-\left(3-|K_S|^2K^{S\,S^*}\right)|W|^2\right].
\label{susypot}
\end{equation}
Here,
$T$ and $S$ are the scalar components of the chiral superfields $\cal T$
and $\cal S$, respectively;    $K_T=\partial_T K$
and
$K_{TT^*}=\partial_T\partial_{T^*} K$ is the
positive definite K\"ahler
metric, and  $D_TW=\partial_TW+K_TW$.

We assume that the matter part of the  potential (\ref{susypot})  has isolated
minima  and the matter field  $T$\ is
responsible for the formation of the wall.
When the potential (\ref{susypot})\ has isolated
supersymmetric minima, {\em i.e\/}.\ when
$\left. D_TW\right|_{T_{1,2}}=0$, there are  extreme walls, which are {\em
static\/}
configurations interpolating between these
minima.

Because of the restricted  form of the K\" ahlerpotential  and
the superpotential
(\ref{kalpot}), the form of the potential (\ref{susypot})\ resembles
closely, although not completely,  the form of the
potential
in Eq.~(\ref{potential})\
with  $\widehat V(\phi)=0$. In other words, a supersymmetric embedding of
dilatonic walls requires a specific type of  potential (\ref{susypot}).
Inside the wall region, such a potential  is  in general not of the simple form
(\ref{potential}). Outside the wall region, however,
Eq.~(\ref{susypot})\
{\em does\/} reduce to the
effective dilaton potential of the type (\ref{dilpot}).
In particular,
in the case of the isolated
supersymmetric minima, {\em i.e\/}.\  $\left. D_TW\right|_{T_{1,2}}=0$,
 the potential (\ref{susypot})\ outside the wall
{\em is\/} of the form  of the effective
dilaton potential (\ref{dilpot})\ with
\begin{equation}
\left. \begin{array}{l}
 f(\phi)= e^{K_{\text{dil}}(S,S^*)}
\left(|K_S|^2K^{S\,S^*}-3\right),\ \
\widehat V=0,\\
V_0(\tau_{1,2})=
\left( e^{K_{\text{matt}}}|W|^2\right)_{T_{1,2}}.
      \end{array}
\right.
\label{factors}
\end{equation}

A particularly interesting case is $N=1$ supergravity theory with a matter
chiral superfield and a  linear supermultiplet\cite{Ferrara}.
In the K\" ahler superspace
formalism the dilaton linear supermultiplet can be
expressed in  terms of a chiral supermultiplet $\cal S$ with the
 K\"ahler potential\cite{Ferrara}:
\begin{equation}
K_{\text{dil}}(S,S^*)=-\alpha\ln(S+S^*).
\label{kald}
\end{equation}
With $S$, the scalar component of $\cal S$, written as
$S=  e^{-2\phi/\sqrt\alpha}+i{\cal A}$, the  potential
in Eq.~(\ref{potential})\ is related to
that of Eq.~(\ref{susypot})\ through:
\begin{equation}
\left. \begin{array}{l}
 f(\phi) =  e^{2\sqrt\alpha\phi},\ \
\widehat{V}=0,\\
V_0=  e^{K_M}\left[|D_TW|^2K^{T\,T^*}-(3-\alpha)|W|^2\right].
       \end{array}
\right.
\label{susypota}
\end{equation}

On the other hand, one is also interested in the  self-dual K\" ahler
potential where $K_{\text{dil}}$ has an extremum for finite $S$.
Such a K\" ahler
potential can be motivated by assuming strong-weak (dilaton)\
coupling invariance of the
theory.

In the following subsections we shall
derive the Bogomol'nyi bound on the energy
density and the  Killing spinor equations
for supersymmetric (extreme)\ domain
wall configurations.

\subsection{Bogomol'nyi bound}
\label{Sub:BogomolnyiBound}

In order to derive the corresponding Killing
spinor equations and the Bogomol'nyi
bounds for dilatonic domain walls,
we use the technique of the generalized
Israel-Nester-Witten form \cite{Nester},
which was originally applied to the study of ordinary
supergravity walls \cite{CGR}.
The results  here are
somewhat more  general than those  for extreme walls
with  an exponential dilaton
coupling\cite{Cvetic,Cvetic2}.\footnote{Analogous procedures
were followed in the
derivation of the   Bogomoln'yni bounds for the mass of
the corresponding
charged  black holes \cite{GH,GP}.}
In addition, a generalization
of  the results  to more than one dilaton field
is straightforward,
as long as the dilatons   have no superpotential
and  a
K\" ahler potential
decoupled from the one of the matter field(s).
Since the extreme domain walls are planar and infinite,
we shall derive the
Bogomol'nyi bound for the  energy per unit area of the
wall.
Note also
that a precise definition of the energy density of the
wall is possible only in the thin wall
approximation, namely, when the ``interior'' and the
``exterior'' regions of the wall
are clearly separated.

We consider a generalized
Nester form\cite{Nester}:
\begin{equation}
N^{\mu \nu} = \bar \epsilon\gamma^{\mu \nu \rho}
\widehat\nabla_{\rho} \epsilon
\label{nester}
\end{equation}
where $\epsilon$ is a  Majorana spinor.
 Here
$\widehat\nabla_{\rho}\epsilon\equiv
\delta_{\epsilon}\psi_{\rho}$  and
$\widehat\nabla_{\rho}=2\nabla_{\rho} +
Q_{\rho}$, where $Q_{\rho}=
i e^{K \over 2}
\left( \Re (W) +\gamma^5 \Im (W)\right)\gamma_{\rho}
 - \gamma^5 \Im (K_{T}\partial_{\rho}T) - \gamma^{5}
\Im (K_{S}\partial_{\rho}S)$ and
 $\nabla_{\mu}\epsilon = (\partial_{\mu}
  + {1\over2}\omega^{ab}_{\mu}\sigma_{ab})\epsilon
$; $\psi_{\rho}$ is the spin $3/2$ gravitino field.
Therefore, the explicit  expression for Nester's form is:\footnote{We use
the conventions: $\gamma^{\mu}=e^{\mu}_{\;\;a}\gamma^{a}$ where
$\gamma^{a}$ are the flat  space--time Dirac matrices satisfying
$\{\gamma^{a},\gamma^{b}\}=2\eta^{ab}$,
$\gamma^5=\gamma^0\gamma^1\gamma^2\gamma^3$; $e^{a}_{\;\;\mu}e^{\mu}_{\;\;b}
= \delta^{a}_{\;\;b}$; $a=0,...3$; $\mu=t,x,y,z$.}
\begin{eqnarray}
N^{\mu\nu}&=&
 \bar\epsilon\gamma^{\mu\nu\rho}
\left[2\nabla_{\rho} + i e^{K \over 2}
\left(\Re (W)+\gamma^{5}\Im (W)\right)\gamma_{\rho}\right.\nonumber \\
 & & \left. {}- \Im(K_{T}\partial_{\rho}T)\gamma^{5} -
 \Im(K_{S}\partial_{\rho}S)\gamma^{5}\right]\epsilon.
\end{eqnarray}
Stokes' theorem  ensures the following relationship:
\begin{equation}
{
\int_{\partial \Sigma} \! N^{\mu \nu} d\Sigma_{\mu \nu}
= 2\int_{\Sigma}\!\nabla_{\nu}N^{\mu \nu} d\Sigma_{\mu}
}
\label{stokes}
\end{equation}
where $\Sigma$ is a space-like hypersurface.

After a lengthy calculation (for details related to the derivation of the
expression below, see Appendices in
Ref.~\cite{CGR}),
the volume integral yields in our case:
\begin{eqnarray}
  2\int_{\Sigma}\!\nabla_{\nu}N^{\mu \nu} d\Sigma_{\mu}
 &=&  \int \!
\biggl[{\overline{\widehat\nabla_{\nu}\epsilon}}\,
\gamma^{\mu\nu\rho}{\hat\nabla_{\rho}\epsilon}
+
K_{T  T^*}\overline{\delta_{\epsilon}\chi} \gamma^{\mu}\delta_{\epsilon}\chi
\nonumber \\
& & {}
+ K_{S
{S}^*} \overline{\delta _{\varepsilon}\eta} \gamma^{\mu}
\delta_{\varepsilon}\eta
+(G^{\mu\nu}-T^{\mu\nu}){\overline{\epsilon}}\gamma_\nu\epsilon
\biggr]d\Sigma_{\mu}
   \ge 0,
\label{volume}
\end{eqnarray}
where $\delta_{\epsilon}\chi$ and
$\delta_{\epsilon}\eta$
are the supersymmetry transformations of fermionic
partners $\chi$ and $\eta$  to  the matter field $T$ and the
dilaton field $S$, respectively; $T^{\mu\nu}$
is the energy-momentum
tensor and $G^{\mu\nu}$ is the Einstein tensor. The first term in
Eq.~(\ref{volume})\ is non-negative, provided
the spinor $\epsilon$ satisfies the (modified)\
Witten condition,
{\em i.e\/}.\
${\mbox{\boldmath{$n$}}}\widehat{\mbox{\boldmath{$\nabla$}}}\epsilon =0$
({\boldmath{$n$}} is the four-vector  normal to
{\boldmath{$\Sigma$}}).
The
K{\"a}hler metric coefficients $K_{T T^*}$ and $K_{S S^*}$
are positive definite,
and thus the second and
the third terms in Eq.~(\ref{volume}) are non-negative
as well. The last term in  Eq.~(\ref{volume})\ is zero
due to Einstein's equations.   Thus, the  integrand in
Eq.~(\ref{volume})
is always non-negative and it is zero  if and only if
the supersymmetry transformations on the gravitino
$\psi_{\rho}$ as well as on
$\chi$ and  $\eta$  vanish, {\em i.e\/}.\
if the
configurations  are supersymmetric.

The surface integral of
Nester's form  in Eq.~(\ref{stokes}) yields the  corresponding Bogomol'nyi
bound for the energy  associated with the  configuration.  We shall
derive the energy  per unit area  (energy density) of the wall   by
evaluating
the corresponding  density of the  surface integral  in Eq.~(\ref{stokes}).
Such a  bound can be derived precisely
only in the thin wall approximation,
because
the region inside the wall must be
clearly separated from the region outside the wall in order
for  its energy density to be well defined.

Within the above assumptions the
space-like  hyper--surface
$\Sigma$  extends  in the $z$-direction only.
The measure is
$d\Sigma_{\mu} = (d\Sigma_{t},0,0,0)$ and
$d\Sigma_t=\sqrt{|g_{tt}g_{zz}|}dz$.
For a thin wall located at
$z=0$ the boundary associated with the surface integral
are then the  two points at $0^+$ and  $0^-$. In addition,  at the location of
the  thin wall,  the metric coefficient $a(0)=0$, and the dilaton
has the  value
$S(0)=S_0$.
Thus, the  corresponding density of
the  surface integral of Nester's form  in Eq.~(\ref{stokes})\
is  of the form:
\begin{equation}
\left.\bar\epsilon_0\gamma^0\epsilon_0\sigma
+\bar\epsilon_0\gamma^{03}
 e^{K/2}\left[\Re(W)+\gamma^5\Im(W)\right]
\epsilon_0\right|_{0^-}^{0^+}.
\label{surface}
\end{equation}
The spinor $\epsilon_0$ is
defined at the boundaries  $z=0^+$ and $z=0^-$ of the wall.
In the first
term we have used the fact
that for the thin wall the magnitude of the spinor
components does not change. The first term  of the surface integral
(\ref{surface})\ of the Nester's form (\ref{nester})\
can then be
identified with the energy density of the wall.
The second term
corresponds to the  topological
charge density $C$ evaluated on  both sides of the wall.
Positivity 
of the volume integral (\ref{volume})\
translates  through Eq.~(\ref{stokes})\
into the corresponding Bogomol'nyi bound
for the energy density of  a thin wall:
\begin{equation}
\sigma \ge
|C|,
\label{localbound}
\end{equation}
which is saturated if and only if
the bosonic background is supersymmetric.

In the following subsection
we shall evaluate  the explicit  phase factors by which
the components  of the
$\epsilon_0$ spinor  change at the wall
boundaries for the case of extreme solutions.
These phase factors
will in turn allow us to
obtain the  explicit form of $\sigma_{\text{ext}}=|C|$.

\subsection{Killing spinor equations}
\label{Sub:KillingSpinorEqs}

We now write down
explicit  Killing spinor equations, {\em i.e\/}.\
$\delta\psi_{\mu} =
\delta\chi =\delta\eta=0$. Killing spinor equations are
satisfied by supersymmetric, static
configurations.
With the metric Ansatz (\ref{metricansatz})\
with $\beta =0$:
\begin{equation}
ds^2= e^{2a(z)}\left(dt^2-dz^2-dx^2-dy^2\right)
\label{extmet}
\end{equation}
and $T(z)$ and $S(z)$ being functions only of $z$, the Killing spinor
equations are of the form:
\begin{mathletters}
\begin{eqnarray}
\delta\chi& = & -\sqrt2\left[ e^{K\over2}K^{T  T^*}
\left(\Re( D_{T}W)+\gamma^5\Im(  D_{T}W)\right)
+ i  e^a\left(\Re (\partial_{z}T )
+ \gamma^5(\partial_{z} T) \right)
 \gamma^{3} \right]\epsilon\\
\delta\eta& = & -\sqrt2 \left[ e^{K\over2}K^{S S^*}
\left(\Re( K_SW)+\gamma^5\Im(K_SW)\right)
+i e^a\left(\Re (\partial_{z}S )
+ \gamma^5(\partial_{z} S) \right)
 \gamma^{3}\right] \epsilon\\
\delta\psi_x&=& \left[- \gamma^{1}\gamma^{3}\partial_{z}a
- i \gamma^{1} e^{(a+{K\over 2})}\left(\Re W+\gamma^5\Im W\right)
\right]\epsilon\\
\delta\psi_y&=& \left[- \gamma^{2}\gamma^{3}\partial_{z}a
- i \gamma^{2} e^{(a+{K\over 2})}\left(\Re W+\gamma^5\Im W\right)
\right]\epsilon \\
\delta\psi_z &=& \left[ 2\partial_{z}
 - i\gamma^{3} e^{(a+{K\over 2})}\left(\Re W
+\gamma^5\Im W\right)
-\gamma^{5}\Im(K_{T}\partial_{z}T+
K_{S}\partial_{z}S)\right]\epsilon\\
\delta\psi_t &=& \left[\gamma^0\gamma^3 \partial_za
 +i \gamma^{0} e^{(a+{K\over 2})}\left(\Re W+\gamma^5\Im
W\right)\right]\epsilon.
\end{eqnarray}
\label{kileq}
\end{mathletters}\\[-1.0em]
We have assumed that  the Majorana spinor $\epsilon=(\epsilon_1,
\epsilon_2,\epsilon_2^*,-\epsilon_1^*)$  does not  depend on
$x^{i}\in\{t,x,y\}$. Note that  in Eqs.~(\ref{kileq})\ the K\" ahler potential
$K=K_{\text{dil}}(S,S^*)+K_{\text{matt}}(T,T^*)$
is separable  and $W=W(T)$,    cf.\ Eq.~(\ref{susylag}).

The vanishing of the above expressions  yields
first-order
differential
equations\footnote{Eqs.~(\ref{kileq})\ equal zero
can be viewed as ``square roots'' of
the corresponding Einstein and
Euler-Lagrange equations;
they provide a particular solution of the equations
of motion which saturate the Bogomol'nyi bound (\ref{localbound}).
The existence  of such
static wall solutions is due to the constrained form of the
matter potential in $N=1$ supergravity theory.
Note also that in the thin
wall approximation one can explicitly solve the  Einstein
equations for $a(z)$  and  the Euler-Lagrange equation for
$S(z)$  outside the wall and then match the solution
for $a(z)$ and $S(z)$ across the wall region.}
(self-dual or
Bogomol'nyi equations)\ for the metric
coefficient $a(z)$,
$T(z)$ and $S(z)$
as well as the constraint on
the spinor $\epsilon$. The field equations are of the form:
\begin{mathletters}
\begin{eqnarray}
0&=&\Im \left(\partial _{z}T{ {D_{T} W}\over W}\right)
\label{boge1}\\
\partial_{z} T &=&\zeta  e^{(a+{K\over 2})}
 |W| K^{T{T^*}}
{{D_{{T^*}} {W}^* }\over{ W^*}}
\label{boge2}\\
\partial_{z}  a &=& \zeta
e^{(a+ {K\over 2})}|W|
\label{boge3}\\
\partial _{z}S &= &-\zeta
e^{(a+{K\over 2})}|W|K^{SS^*}K_{S^*}.
\label{boge4}
\end{eqnarray}
\label{boge}
\end{mathletters}\\[-1.0em]
Here  $\zeta=\pm 1$ and
it
can change sign only when $W$ crosses zero.
There is another constraint
on the  ``field geodesic'' motion of the
dilaton field, namely  $\Im(K_{
S}\partial_z S)=0$.
However,
by multiplying Eq.~(\ref{boge4})\
by $K_{  S}$,
this constraint is
seen to be automatically satisfied.
In this case the right-hand side
of the equation is real,
since $K^{SS^*}> 0$ is real and
$K_{S^*} =(K_{S})^*$.

Eqs.~(\ref{boge1})\ and (\ref{boge2})\
describe the evolution of the
matter field $T = T(z)$ with $z$.
By now,
Eq.~(\ref{boge1})\ is
a familiar ``field
geodesic'' equation,
which  determines
the path of the complex scalar field $T$  in the complex
plane  between the two
minima $T_1$ and $T_2$ of the matter potential. It can
always be satisfied for
Type~I walls (those with $W(T_1)=0$).
Eq.~(\ref{boge2})\ governs the change of
the $T$ field with coordinate $z$ along this path
(analogous to  the field
$\tau$ in  Section~\ref{Sect:ThinWall}).

Eqs.~(\ref{boge3})\ and (\ref{boge4})\
determine the evolution of the
metric coefficient $a(z)$ and the
complex  field $S$.
These two equations
imply another
interesting relation between  the  dilaton K\"
ahler potential $K_{\text{dil}}(S,S^*)$ and  $a(z)$:
\begin{equation}
2K^{SS^*}|K_{S}|^2\partial_za+\partial_zK_{\text{dil}} =0.
\label{ap}
\end{equation}

In addition, the Killing spinor
equations (\ref{kileq})\   impose
a constraint on the phase of the Majorana spinor. Namely,
the solution for the Killing spinor component  is of the form:
\begin{equation}
\epsilon_1=  e^{i\theta}\epsilon_2^*={\cal C} e^{(a+i\theta)\over 2},
\label{killsp}
\end{equation}
where the phase $\theta(z)$ satisfies the following equation:
\begin{equation}
\partial_z\theta=-{\Im}(K_T\partial_zT).
\label{thet}
\end{equation}
The constant ${\cal C}$ can be set to 1/2 for
the Majorana spinors normalized as
$\epsilon^\dagger\epsilon=1$. The constraint (\ref{killsp})\
on the Killing
spinor $\epsilon$ in turn
implies that the extreme configurations preserve
``$N={1\over 2}$''
of the original $N=1$ supersymmetry.

The energy  density of the  wall is determined by setting
Eq.~(\ref{surface})\ to zero. With the explicit
form for the Killing spinor
components (\ref{killsp}),
Eq.~(\ref{surface})\ yields:
\begin{eqnarray}
\sigma_{\text{ext}}=|C| & = &
2 \left|\left(\zeta e^{K\over 2}W\right)_{z=0^+}-
\left(\zeta e^{K\over
2}W\right)_{z=0^-}\right| \nonumber\\
& = & 2e^{{K_{\text{dil}}(S_0,S_0^*)\over 2}}\left|\left(\zeta
e^{{K_{\text{matt}}} \over 2}
W\right)_{z=0^+}-\left(\zeta  e^{{K_{\text{matt}}} \over 2}
W\right)_{z=0^-}\right|.
\label{sig}
\end{eqnarray}
Here the subscript $z=0^{\pm}$ refers to  either  side of the wall.
Without loss of generality
we have normalized $a(0)=0$ and set $S(0)=S_0$.

\paragraph*{Classification of extreme domain wall solutions:}
Solutions to the Bogomol'nyi equations (\ref{boge})\
fall into three types,
depending  on whether
$W(T)$ crosses zero
or not
along the wall
trajectory:
\begin{itemize}
\item[{}]{\bf Type I} walls correspond to those
where on one side of the wall, say for
$z>0$, $W(T_1)=0$. In this case the
energy density of the wall is of the form:
$\sigma_{\text{ext}}=2 \left| e^{K\over
2}W\right|_{z=0^-}$.
Note that in this
case the side of the wall with $z>0$
corresponds to the Minkowski space--time with a constant $S$.
\item[{}]{\bf Type II} walls
correspond to the walls with $W(T)$ crossing zero
somewhere along the wall trajectory.
In this case  $\zeta$ changes sign at
$W=0$. The energy density of the wall is specified by:
$\sigma_{\text{ext}}=2 \left|  e^{K\over
2}W\right|_{z=0^+}+\left| e^{K\over 2}W\right|_{z=0^-}$.
Reflection symmetric
walls  fall into this class.
\item[{}]{\bf Type III} walls correspond to the walls where $W(T)\ne 0$
everywhere in
the domain wall background. In this  $\zeta$ does not change
sign. The energy density of such walls is:
$\sigma_{\text{ext}}=2 \left|\left|  e^{K\over
2}W\right|_{z=0^+}-\left|  e^{K\over 2}W\right|_{z=0^-}\right|$.
\end{itemize}

In the following  section we shall  concentrate
on the explicit form of the extreme
solutions  using some of the formalism
spelled out in the previous  two sections.

\section{Explicit form of extreme solutions}
\label{Sect:Explicit}

This section concentrates on the Einstein-dilaton system
outside the extreme  domain wall region.
The  explicit form of  the extreme solutions
in the thin wall
approximation\footnote{Explicit numerical solutions of
Eqs.~(\ref{boge})\
for  a wall of any thickness
have  the same qualitative features outside the
wall region.}
are presented.

We shall first recapitulate  results for  extreme  walls
with $K_{\text{dil}}=-\alpha\ln (S+S^*)$ \cite{Cvetic2}.
Then we shall study extreme domain walls with self-dual
$K_{\text{dil}}$,  {\em i.e\/}.\
$\left. K_{S}\right|
{\mbox{\raisebox{-1.0ex}{{\tiny{$\!S'\!$}}}}}
=0$
for some $S'$. An example
of the latter class
corresponds to a solution of a
theory  with  a strong-weak (dilaton)\
coupling symmetry, {\em i.e.\/},
$SL(2,{\bf Z})$ invariance of
the dilaton coupling.

Subsection~\ref{Sub:ExExp}
presents the extreme solutions in theories with
exponential dilaton coupling. Their physical properties such as
the Hawking
temperature associated
with the horizons and the
gravitational mass of the singularities are
also discussed.
In Subsection~\ref{Sub:Correspondence}
the correspondence with cosmological
models is used to find the necessary and sufficient condition
for the existence of horizons.
Subsection~\ref{Sub:Self-dual}
comments on the
self-dual extreme solutions.

\subsection{Extreme solutions for exponential dilaton coupling}
\label{Sub:ExExp}

Let us first consider the gravitational properties of extreme
super\-symmetric domain walls
with $K_{\text{dil}}=-\alpha\ln(S+S^*)$, which has been
worked out in Refs.~\cite{Cvetic,Cvetic2}.
The  case  with
$\alpha = 2$ had earlier been found in the
guise of a static plane-symmetric space--time
with a
conformally coupled
scalar field \cite{VaidyaSom,AcciolyVaidyaSom}. In
Ref.~\cite{GS} this space--time
was shown to be induced by a domain wall.
Here we shall
recapitulate the results and
focus on the relationship with cosmological
solutions.

With
the notation $S=  e^{-2\phi/\sqrt\alpha}+i{\cal A}'$,
the Bogomol'nyi equations of motion are of
the type:
\begin{mathletters}
\begin{eqnarray}
0&=&\Im\left(\partial _{z}T {{D_{T} W}\over W}\right)
\label{bogep1}\\
\partial_{z} T& = &-\zeta2^{-{\alpha\over 2}}
e^{(a+\sqrt\alpha\phi)} e^{K_{\text{matt}}\over
2} |W| K^{T {T}^*}
{{D_{{T}^*}{W}^*}\over W^*}
\label{bogep2}\\
\partial_{z} a& = &\zeta 2^{-{\alpha\over 2}} e^{(a+\sqrt\phi)}
 e^{K_{\text{matt}}\over 2}|W|
\label{bogep3}\\
\partial _{z}\phi& =& - \zeta\sqrt\alpha2^{-{\alpha\over 2}}
 e^{(a+\sqrt\alpha \phi)}
 e^{K_{\text{matt}}\over 2}|W|,
\label{bogep4}
\end{eqnarray}
\label{bogep}
\end{mathletters}\\[-1.0em]
and the axion ${\cal A}'$, the imaginary part of $S$,
is constant. The energy density of the
wall is of the
form:
\begin{equation}
\sigma_{\text{ext}}=  2^{1-{\alpha \over 2}}
e^{\sqrt\alpha
\phi_{0}}\left| \left(e^{{K_{\text{matt}}} \over 2}
W\right)_{z=0^+} \pm \left(e^{{K_{\text{matt}}} \over 2}
W\right)_{z=0^-}\right|\equiv 2(\chi_1\pm\chi_2) \label{sigp}
\end{equation}
where we have chosen the boundary condition for
$a(0)=0$ and $\phi(0)=\phi_0$.
The $+$ and $-$ signs correspond to the Type~II and Type~III walls,
respectively.
It is understood that the coordinates are chosen so that
$\chi_2\leq \chi_1$.  With the  choice
$\chi_{1,2}=2|e^{K/2}W|_{z=0^\mp}$, the solution of Eqs.~(\ref{boge})\
and
(\ref{bogep})\ are satisfied with the following  choice for $\zeta$:
on the $z<0$ side,  $\zeta=1$, while on the
$z>0$, $\zeta=-1$ for
Type~II walls and   $\zeta= 1$ for Type~III walls.

The solution of the
Bogomol'nyi equations (\ref{bogep})\
outside the wall region, {\em i.e\/}.\ when
$\partial_z T\sim 0$,
are the same as those of
 the second-order equations (\ref{fieldeqs})\
with  $\beta=0$ and  the effective
dilaton potential  of the type of Eq.~(\ref{dilpot}),
where
\begin{equation}
f(\phi)= e^{2\sqrt{\alpha}\phi}, \ \
V_0(\tau_{1,2})=
-(3-\alpha)2^{-\alpha}\left.\left(
e^{K_{\text{matt}}}|W|^2\right)\right|_{T_{1,2}}, \ \
\widehat V (\phi)=0
\label{parexp}
\end{equation}
on either side of the wall.
The  value of parameters $\chi_{1,2}$  in
Eq.~(\ref{sigp}) on either
side of the wall is related to $ V_0(\tau_{1,2})$ in the following way
\begin{equation}
\chi_{1,2}\equiv  2^{-{\alpha\over 2}} e^{\sqrt\alpha
\phi_0}\left( e^{{K_{\text{matt}}\over 2}}|W|\right)|_{T_{1,2}}= e^{\sqrt\alpha
\phi_0}\sqrt{-V_0(\tau_{1,2})/(3-\alpha)}.
\label{chidetermined}
\end{equation}
Note that $\alpha=3$
corresponds to  the point where $V_0$ changes sign.

The explicit
solution on either side of the wall is of the form
\begin{mathletters}
\begin{equation}
 a = \left\{
     \begin{array}{rl}
 \chi_1 z, &   \;
 \ \ z<0\\
\mp \chi_2 z,     &   \;
\ \ z>0
     \end{array}
     \right.
\label{stringextreme}
\end{equation}
if $\alpha=1$ and
\begin{equation}
a= \left\{
     \begin{array}{ll}
(\alpha-1)^{-1}\,\ln [1+\chi_1 (\alpha-1 )z ], &
\ \  z<0\\
(\alpha-1)^{-1}\,\ln [1\mp\chi_2 (\alpha-1 )z ], &
\ \  z>0
     \end{array}
     \right.
\label{extremesolution}
\end{equation}
\label{extremesolutions}
\end{mathletters}\\[-1.0em]
if $\alpha\neq 1$.
The upper and lower signs  of the solutions
(\ref{extremesolutions})\
correspond to the Type~II and
Type~III solutions, respectively.
Type~I corresponds to the special case with $\chi_2=0$, {\em i.e\/}.\
those are
solutions with Minkowski space--time  ($a=0$)\
 and a  constant dilaton on  the $z>0$
side of the wall.

Note that Eqs.~(\ref{bogep3})\ and (\ref{bogep4})\ imply that
\begin{equation}
\phi = -\sqrt{\alpha}a
\end{equation}
everywhere in the domain wall background.  Consequently,
these solutions are represented by  straight lines in the
$(a',\phi')$ phase diagram.

For $\alpha>1$ the domain walls have a naked (planar)\
singularity  at
$z=-1/[\chi_1(\alpha-1)]$  and for Type~II
walls at $z=1/[\chi_2(\alpha-1)]$
as well.  For $\alpha \le 1$
the singularity becomes null, {\em i.e\/}.\ it
occurs at $z=-\infty$ and for
Type~II walls at $z=\infty$ as well. Note that
for $\alpha< 1$ Type~III walls have a coordinate singularity  at
 $z=1/[\chi_2(1-\alpha)]$.
Thus, extreme walls with the
``stringy'' coupling $\alpha=1$ act as a window
between the extreme dilatonic walls
with naked singularities and those with
singularities covered by a horizon.

\begin{figure}[htbp]
{\vspace{-1.7cm}}
\if\FiguresAvailable\NO
    {\vspace{5cm}}
\else
\centerline{
{\hspace{-1.5cm}}
\psfig{figure=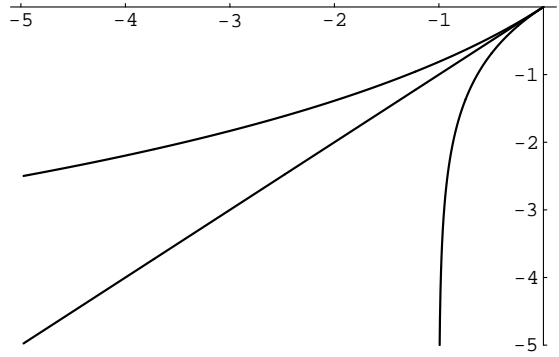,height=10.1cm,silent=}
\vspace{-2.2cm}}
\fi
{\caption{\baselineskip 3 ex
$a$ versus $z$ in units of $\chi$ with
for extreme solutions with
$\alpha = 0.5$, $\alpha=1$, and $\alpha=2$, respectively.
Solutions with $\alpha>1$ collapse to a naked singularity
at a finite value of $z$.
\label{ext_a}}}
\end{figure}

The evolution of $a$ for different values of $\alpha$
is plotted in Fig.~\ref{ext_a}.

\paragraph*{Temperature and gravitational mass per area:}
Static domain wall configurations
with   space--time singularities are only possible
if there is an exact cancellation of the contributions to the
gravitational mass coming from the wall, the dilaton field,
and from the singularity.
Let us consider the case of  an
extreme Type~I wall, {\em i.e.\/}
a
static dilatonic wall with a non-zero vacuum energy on
one side (say, $z<0$  and $\chi_1\ne 0$)\
and a Minkowski space on the other
($z>0$ and $\chi_2=0$).

We shall employ the
concept of gravitational mass per area, $\Sigma$, as derived in
Ref.~\cite{CGS}. It can be written in the form:
\begin{equation}
\Sigma(z)=\frac{\int_{-\infty}^{z}\sqrt{-g^{(4)}} dz'
\left(T^{t}_{\;\;t}-T^{z}_{\;\;z}-T^{x}_{\;\;x}-T^{y}_{\;\;y}\right)
\int dx dy}{\sqrt{g^{(2)}} \int dx dy}.
\end{equation}

This is a plane-symmetric version of
Tolman's \cite{Tolman} mass formula.
The contribution from all sources outside the horizon
(or the naked singularity)\
can be
expressed as
\begin{equation}
\Sigma (\infty) = \left.2 a'\right|
{\mbox{\raisebox{-1.2ex}{{\tiny{$\!\infty$}}}}}
 - \left.2 a'\right|
{\mbox{\raisebox{-1.2ex}{{$\!$\tiny{horizon}}}}}.
\end{equation}
On the Minkowski side $a'\equiv da/dz=0$, and so $\left.a'\right|
{\mbox{\raisebox{-1.2ex}{{\tiny{$\!\infty$}}}}}=0$.
Hence, $\Sigma$ is determined by the value of $a'$ at the horizon.
For $\alpha<1$,
the gravitational mass per area
outside the horizon vanishes.
This means that there is no mass beyond the horizon.
The
Hawking temperature
associated
with this horizon also vanishes.

For $\alpha=1$, the gravitational mass per area
from sources
outside the
horizon is negative: $\Sigma (\infty)= -2\chi_1$. In order
to have a vanishing total mass,
the mass of the singularity must be
$\Sigma_{\text{singularity}}  = 2\chi_1$, but
a proper mathematical description would
require a distribution-valued
metric.
We shall not pursue this issue further here,
but we note that a
source at the singularity of the
Schwarzschild metric
has recently been identified in terms of a
distributional energy-momentum tensor
\cite{BalasinNachgebauer}.
The Hawking temperature for the extreme
Type~I wall with $\alpha=1$ is finite:
$T=\chi_1/2\pi$\cite{Cvetic2}.

For $\alpha>1$, the mass per area outside the singularity
is negative and infinite. Accordingly, there must be an
infinite positive mass in the singularity, which  is then
a singularity even in
a distributional sense. This naked
singularity also has an infinite Hawking temperature.

\subsection{Correspondence with cosmological solutions}
\label{Sub:Correspondence}

The equivalence  of  the domain wall solutions on either side of the
wall  and a  class of cosmological solutions implies that
we are able to relate the
above solutions to known solutions of inflationary cosmology
with exponential potentials \cite{Barrow}, which were
later
generalized to higher-dimensional
FLRW cosmologies \cite{BurdBarrow}.
Properties  of general (extreme  and non-extreme)\
scalar field cosmological models and their
corresponding phase diagrams were studied  in
Ref.~\cite{Hall}.\footnote{Note that
in the cosmological picture the extreme Type~I  vacuum domain wall
becomes a flat inflationary universe where the inflaton
(the wall-forming scalar field in the original picture)\
rolls down the inflation potential with just the right speed so that
it
stops at a local maximum
corresponding to a vanishing cosmological
constant. At this point the universe also stops expanding.}

Note that after the substitution
$z\rightarrow \eta$ and $V_0\rightarrow -V_{0 \text{c}}$
the Type~II solutions
and Type~III solutions (on the $z>0$ side)\ correspond to
to contracting and expanding
cosmological solutions,
respectively. For cosmological models
$\chi_{1,2}\equiv\sqrt{V_{0\text{c}}/(3-\alpha)}$.
The value $\alpha=3$ corresponds to the point
where $V_{\text{c}0}$ changes sign
from positive (for $\alpha<3$) to negative (for $\alpha>3$).

Since the  extreme solutions are characterized by
$\phi = -\sqrt{\alpha}a
$ they  are represented by  straight lines in the
$(\dot a,\dot\phi)$ phase diagram\cite{Hall}.

\paragraph*{Cosmological horizons and domain wall event horizons:}
We would now like to relate the nature of the cosmological horizons to the
event horizons in the domain wall background.
If we write the cosmological line element in the standard form
\begin{equation}
ds^2 = d\tau^2 - R^2(\tau)\, \left(\frac{dr^2}{1+\beta^2 r^2} + r^2
d\Omega_{2}^2\right),
\end{equation}
then the {\em convergence\/} of the integral
\begin{equation}
I=\int^{\tau_{1}}_{\tau_{0}}\! \frac{d\tau }{R(\tau)}
=\int^{\eta_{1}}_{\eta_{0}}\! d\eta
\end{equation}
in the limit $\tau_{1}\rightarrow \tau_{\text{max}}$
is a necessary and sufficient condition for the existence of
a cosmological event horizon \cite{Weinberg}.
Note, however, that
the complex rotation to the domain wall space--time interchanges
a space dimension with the time
dimension.
Because of this,
the sufficient and necessary condition for having
an event horizon in the domain wall space--time is
that
\begin{equation}
I=\int^{\eta_{\text{max}}}_{\eta_{0}}d\eta = \eta_{\text{max}}-\eta_{0}
\end{equation}
{\em diverges\/}.
In other words, if there is a singularity at finite $\eta$, then
this singularity is naked.
\paragraph*{Equation of state:}
For $0\leq \alpha\leq 3$,
the equation of state
is given by
\begin{equation}
\gamma = \frac{2\alpha}{3}.
\end{equation}
The ``stringy'' value, $\alpha=1$, is therefore the dividing line
between the solutions corresponding to attractive and repulsive
equations of state in the cosmological picture.
In the domain wall system, this is the dividing line of domain walls
with naked singularities ($\alpha >1$), and domain walls
with
the singularity hidden behind a horizon ($\alpha < 1$).

\begin{figure}[htbp]
{\vspace{-1.7cm}}
\if\FiguresAvailable\NO
    {\vspace{5cm}}
\else
\centerline{
{\hspace{-1.5cm}}
\psfig{figure=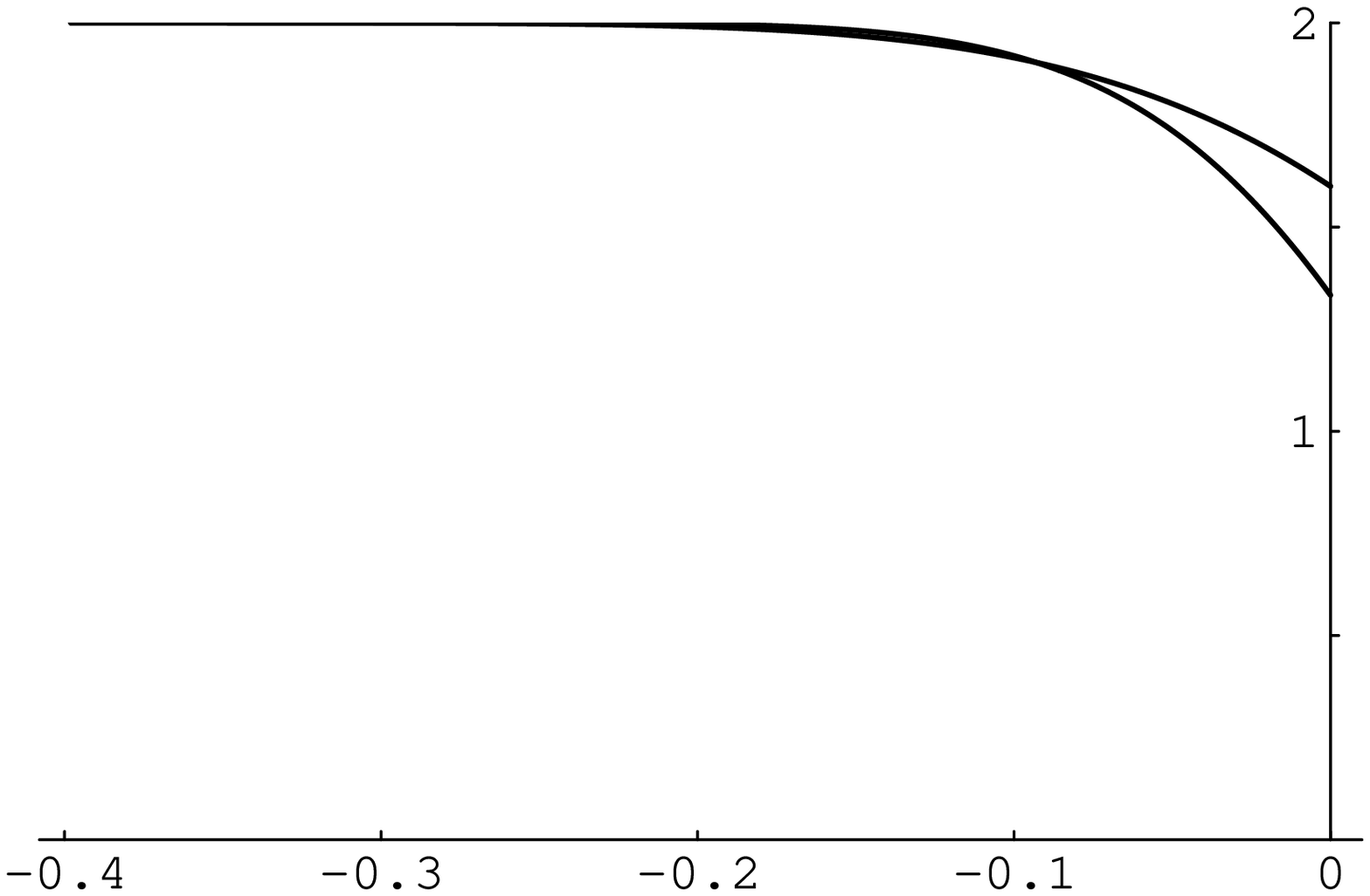,height=10.1cm,silent=}
\vspace{-2.2cm}}
\fi
{\caption{\baselineskip 3 ex
$\gamma$ versus $z$ in units of $\chi$ for extreme solutions with
with $\alpha = 4$ (upper curve)\ and
$\alpha = 6$.
For $0\leq \alpha\leq 3$, the equations of state are
straight lines $\gamma = 2\alpha/3$.
\label{ext_gamma}}}
\end{figure}

For $\alpha>3$ the equation of state is time-dependent
(see Fig.~\ref{ext_gamma}).
It approaches $\gamma=2$ near the singularity.


\subsection{Self-dual dilaton coupling}
\label{Sub:Self-dual}

We would also like to address extreme domain wall solutions  corresponding to
the self-dual dilaton K\" ahler potential $K_{\text{dil}}$.
Namely, $K_{\text{dil}}$  has an extremum
($K_S|_{S'}\equiv\partial K_{\text{dil}}/\partial S|_{S'}=0$) for
some  finite $S =S'$.
The hope is that outside the wall region the dilaton
would reach  the point $S=S'$, and from then on it would remain constant. Such
solutions would in turn reduce to singularity-free space--times of vacuum
domain walls. We shall see that such  a class of extreme solutions is not
dynamically stable within $N=1$ supergravity theory.

In particular, we shall show that if at $z\sim 0$,
$S(0)= S'+\Delta(0)$, with $\Delta(0)$ being an
infinitesimal perturbation from the
self-dual point, $S=S'$, then
$\Delta(z)$ grows indefinitely as $z\rightarrow
-\infty$ and thus the solution with a
constant dilaton outside the wall region is
not dynamically stable.
On  the $z<0$   side of the wall,
$\Delta (z)$ (and likewise for $\Delta^*$)\
is a solution of the following equation:
\begin{equation}
\Delta''-(a'+\zeta\chi_1e^a)\Delta' +
\chi_1^2\left(1-|K_{SS}K^{SS^*}|^2\right)e^{2a}\Delta= 0 ,
\label{delp}
\end{equation}
where  $\chi_1=(
e^{K/2}W)|_{S',T_1}$
and $\zeta=1$ on  the $z<0$ side  of the wall (see
conventions spelled out  after Eq.~(\ref{sigp})).
Eq.~(\ref{delp})\
is obtained
by  expanding   $S(z)=S'+\Delta(z)$
in Eq.~(\ref{boge4}).
In the above expansion one has used
$K_S|_{S'}=0$ and $K^{SS^*}>0$.

In Eq.~(\ref{delp})\
it  suffices to use
the   metric coefficient $a(z)$, which is  determined as a
zeroth-order solution of
Eq.~(\ref{boge3}). Namely, in
Eq.~(\ref{boge})\
one sets $S=S'$, and thus  $a(z)=-\ln(1-\chi_1z)$ is the
solution for the anti-de~Sitter
space--time with the cosmological constant
$\Lambda=-3\chi_1^2$  (see Eq.~(\ref{extremesolution})\
with $\alpha=0$).
Eq.~(\ref{delp}) for
$\Delta$ can then be rewritten  as:
\begin{equation}
{{d^2\Delta}\over {dy^2}}-2{{d\Delta}\over{dy}}
+ (1-|K_{SS}K^{SS^*}|^2)\Delta= 0,
\label{del}
\end{equation}
where  $y\equiv\ln(1-\chi_1 z)$. Consequently, the
general form of the solution is:
\begin{equation}
\Delta(z)={\cal A}_1(1-\chi_1z)^{(1+|K_{SS}K^{SS^*}|)}+
{\cal A}_2(1-\chi_1z)^{(1-|K_{SS}K^{SS^*}|)},
\label{delsol}
\end{equation}
where ${\cal A}_{1}$ and ${\cal{A}}_{2}$
are
complex constants determined by the initial conditions of
$\Delta$. The solution (\ref{delsol})\
for
$\Delta (z)$
grows indefinitely as $z\rightarrow -\infty$, thus  implying
instability of the
constant dilaton  ($S=S'$)\
solution.
Therefore, on the  $z<0$
side of  extreme walls {\em the constant dilaton
solution is
always dynamically unstable\/};
any small deviation  in the boundary condition $S(z)=S'$
near $z\sim 0$   leads to
a space--time with a varying dilaton
field where  one encounters
planar singularities, in general. Whether such
singularities are naked or not depends on
the nature of the self-dual K\" ahler potential.

On the  $z>0$ side of  the
Type~II wall the constant dilaton solution is also
unstable;  $\Delta(z)$
satisfies  the same type of
equation as Eq.~(\ref{del}),
but where now $y\equiv\ln(1+\chi_2z)$ and
$\chi_2=(
e^{K/2}W)|_{S', T_2}$. Thus, as $z\rightarrow \infty$,
$|\Delta(z)|\rightarrow\infty$ and the solution is unstable.
On the other hand, on the $z>0$ side  of Type~III walls
 $\Delta(z)$
satisfies  Eq.~(\ref{del}) with
$y\equiv\ln(1-\chi_2z)$. In this case  a
general form of the solution is of the type:
\begin{equation}
\Delta(z)={\cal B}_1(1-\chi_2z)^{(1+|K_{SS}K^{SS^*}|)}+
{\cal B}_2(1-\chi_2z)^{(1-|K_{SS}K^{SS^*}|)} ,
\label{delsolp}
\end{equation}
where ${\cal{B}}_{1}$ and ${\cal{B}}_{2}$ are complex constants
determined by the initial conditions of
$\Delta$. Thus, for $|K_{SS}|>K_{SS^*}$, the
constant dilaton solution
on the $z>0$ side of the Type~III wall is  unstable; however, for
 $|K_{SS}|\le K_{SS^*}$,
the solution is {\em stable\/}. In the latter case
the space--time reduces  to the anti-de~Sitter
space--time  where a finite value of
$z\sim 1/\chi_2$  corresponds to the time-like
boundary of space--time.

In conclusion,
in $N=1$ supergravity theory with a self-dual K\" ahler
potential,  extreme
solutions corresponding to the constant dilaton, $S=S'$ (where
$\left. K_S\right|
{\mbox{\raisebox{-1.0ex}{{\tiny{$\!S'\!$}}}}}
=0$),
are always  dynamically unstable (at least on one side of the
wall).

\section{Non- and ultra-extreme solutions}
\label{Sect:Non/Ultra}
In this section we shall analyse
the non-extreme and ultra-extreme
solutions. These are
solutions that are not supersymmetric. They
correspond to the domain wall backgrounds  with moving wall boundaries.
Unlike extreme solutions, which have
supersymmetric embeddings and
where solutions can be given
for any thickness of the wall,
we have only obtained non- and
ultra-extreme solutions
in the thin wall approximation.
We thus employ the formalism spelled
out in Section~\ref{Sect:ThinWall}.

In the following subsections we shall
address the non-extreme solutions for
the exponential dilaton potential,
as well as the case with the self-dual
dilaton potential. We
shall also add a mass term  for the dilaton
field.  Since we have  found only
numerical solutions for non-extreme  walls,
we shall confine
the analysis to the non-extreme Type~I walls (with
Minkowski space--time on one side of the wall)\
and reflection symmetric
solutions, because only in this case
the boundary conditions  on either
side of the wall can be specified uniquely.
In subsection~\ref{Sub:M4-dil} the boundary conditions
for the non- and ultra-extreme Type~I walls
are written down explicitly, and the field equations are reduced
to a first-order system. The results of the numerical integrations
are presented and discussed. Subsection~\ref{Sub:Reflection}
contains solutions for the reflection-symmetric cases,
and in Subsection~\ref{Sub:Self-interaction}
the effects of dilaton self-interactions are studied.
Finally, in Subsection~\ref{Sub:Self-Dual}
it is pointed out that
the singularity-free self-dual
dilatonic domain walls are dynamically unstable.

\subsection{Walls with Minkowski space--time on one side}
\label{Sub:M4-dil}

We now consider the
case where the
dilaton
potential outside the wall region  has the  form
specified in Eq.~(\ref{dilpot})\
with $f= e^{2\sqrt\alpha \phi}$
and $\widehat V(\phi)=0$.
Let the wall be non- or ultra-extreme
with a non-vanishing $V_{0}$ and a running dilaton on one side
($z<0$)\ and a
Minkowski space with $V_{0}=0$ and
a constant dilaton on the other side
($z>0$).
According to the results of
Section~\ref{Sub:FieldAnsatze}, the boundary conditions
are
\begin{equation}
\left. \begin{array}{ll}
 \left.\phi'\right|
{\mbox{\raisebox{-1.2ex}{{\tiny{$\!0^{-}$}}}}}
 = -\frac{1}{2}\sqrt{\alpha}\sigma,\ \ &
\left.a'\right|
{\mbox{\raisebox{-1.2ex}{{\tiny{$\!0^{-}$}}}}}
 =  \frac{1}{2}\sigma -\beta \\
\left.\phi'\right|
{\mbox{\raisebox{-1.2ex}{{\tiny{$\!0^{+}$}}}}}
 = 0\, \ \ &
\left.a'\right|
{\mbox{\raisebox{-1.2ex}{{\tiny{$\!0^{+}$}}}}}
 = -\beta .
       \end{array}
\right.
\label{nonultraboundaryconditions}
\end{equation}
$\beta<0$ and $\beta>0$
represent an ultra-extreme and a non-extreme wall, respectively.  Without loss
of generality  we have also chosen $\phi|
{\mbox{\raisebox{-1.2ex}{{\tiny{$\!0$}}}}}=0$ and normalized the metric
coefficient $a|
{\mbox{\raisebox{-1.2ex}{{\tiny{$\!0$}}}}}=0$. The choice $\phi|
{\mbox{\raisebox{-1.2ex}{{\tiny{$\!0$}}}}}=\phi_0\ne 0$ would correspond to
the rescaling  $V_0\rightarrow {\rm e}^{2\sqrt\alpha\phi_0}V_0$.

At the boundary $z=0^{-}$,
Eq.~(\ref{fieldeqs3})\ gives
\begin{equation}
V_{0}- 3\beta\sigma + (3-\alpha) \left(\frac{\sigma}{2}\right)^2 =0.
\end{equation}
For $\alpha\neq 3$ the above equation reduces to
\begin{equation}
\sigma = \frac{2}{3-\alpha}\left[ \sqrt{ 9 \beta^2 +
(3-\alpha)^2\chi_1^2}+3\beta \right],
\label{walldensity}
\end{equation}
where we have used the fact that
$\chi_1=\sqrt{-V_0(\tau_1)/(3-\alpha)}$, as found in Eq.~(\ref{chidetermined}).
In the case $\alpha=0$, $\beta \neq 0$,
and $\chi_1=\sqrt{|V_0|/3}$,
one recovers the result from the
non- and ultra-extreme anti-de~Sitter--Minkowski space--time
walls  without a dilaton,
and if $\alpha = 0$, $\beta>0$, and $\chi_1=0$,
one recovers the
dilaton-free
non-extreme Minkowski-Minkowski walls \cite{Vilenkin,CGS}.

For $\alpha=3$, one finds
\begin{equation}
V_{0}=3\beta\sigma,
\end{equation}
which indicates that in the non-supersymmetric
case with
$\alpha=3$, the potential
itself gets a non-zero value.

\begin{figure}[htbp]
{\vspace{-1.7cm}}
\if\FiguresAvailable\NO
    {\vspace{5cm}}
\else
\centerline{
{\hspace{-1.5cm}}
\psfig{figure=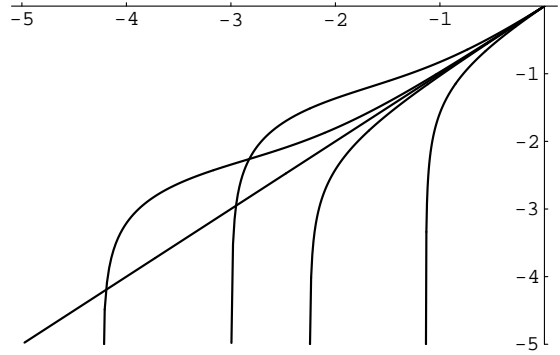,height=10.1cm,silent=}
\vspace{-2.2cm}}
\fi
{\caption{\baselineskip 3 ex
$a$ versus $z$ in units of $\chi$ with $\alpha = 1$ for
different values of $\beta$.
Starting from the left at the bottom of the figure where
$a=-5$, the curves correspond
to $\beta=0$, $\beta=-0.01$, $\beta=-0.1$, $\beta=0.01$, and
$\beta=0.1$.
\label{non_a}}}
\end{figure}

\begin{figure}[htbp]
{\vspace{-1.7cm}}
\if\FiguresAvailable\NO
    {\vspace{5cm}}
\else
\centerline{
{\hspace{-1.5cm}}
\psfig{figure=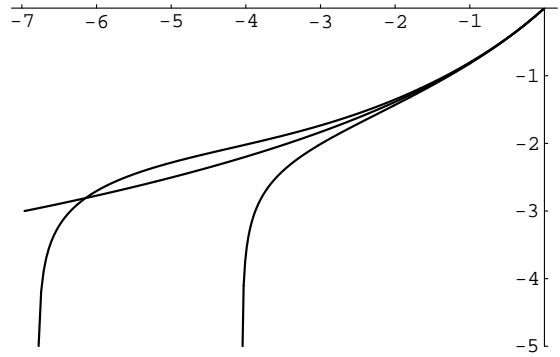,height=10.1cm,silent=}
\vspace{-2.2cm}}
\fi
{\caption{\baselineskip 3 ex
$a$ versus $z$ in units of $\chi$ with $\alpha = 1/2$.
The curve starting in the middle
corresponds to the extreme solution. The non-extreme
case with
$\beta=0.01$ becomes singular shortly after $z = -4$.
The third curve corresponds to
the ultra-extreme
case with $\beta=-0.01$. It also ends in a singularity.
\label{non_a2}}}
\end{figure}

In order to integrate the field equations numerically
we define a new $\tilde{z}$-coordinate  by
\begin{equation}
\tilde z\equiv \chi_1 z.
\end{equation}
Let us also define
\begin{equation}
\left. \begin{array}{ll}
B\equiv a',\ \ &
P\equiv \phi'. \\
\tilde\sigma \equiv \chi_1^{-1}\sigma,\ \ &
\tilde\beta \equiv \chi_1^{-1}\beta,
       \end{array}
\right.
\end{equation}
where a prime now stands for a derivative with respect to the
dimensionless coordinate $\tilde z$.
Here, $\sigma$
is as given in Eq.~(\ref{walldensity}).
Then the field equations (\ref{fieldeqs})\ can be written as
the first-order system
\begin{equation}
\left.
\begin{array}{ll}
a' = B, \ \ &
B' =  e^{2 a + 2 \sqrt{\alpha}\, \phi} (3-\alpha)/3 - 2 P^2/3,\\
\phi' = P, \ \ &
P' = -2 B P - \sqrt{\alpha} \,  e^{2 a + 2 \sqrt{\alpha}\,\phi}
(3-\alpha).
\end{array}
\right.
\end{equation}
The domain wall boundary conditions
(\ref{nonultraboundaryconditions})\ then imply the following
initial conditions for this dynamical system
\begin{equation}
\begin{array}{ll}
\left.a\right|
{\mbox{\raisebox{-1.2ex}{{\tiny{$\!0^{-}$}}}}}
=0, \ \ &
\left.B\right|
{\mbox{\raisebox{-1.2ex}{{\tiny{$\!0^{-}$}}}}}
=\frac{1}{2}\tilde\sigma-\tilde\beta,\\
\left.\phi\right|
{\mbox{\raisebox{-1.2ex}{{\tiny{$\!0^{-}$}}}}}
=0, \ \ &
\left.P\right|
{\mbox{\raisebox{-1.2ex}{{\tiny{$\!0^{-}$}}}}}
=-\frac{1}{2}\sqrt{\alpha}\, \tilde\sigma .
\end{array}
\label{boundarydynamical}
\end{equation}
The  conformal factor goes to zero faster than in the extreme
space--times {\em both\/} in the non- and
ultra-extreme cases (See Figs.~\ref{non_a} and
\ref{non_a2}). This is the case for {\em all\/}
values of $\alpha$ as long as $\beta\ne 0$.
As illustrated in Fig.~\ref{non_a},
when  $|\beta|$ is increased, the  conformal factor decreases even faster.
{\em Such domain walls  thus always exhibit naked singularities\/}.

\begin{figure}[htbp]
{\vspace{-1.7cm}}
\if\FiguresAvailable\NO
    {\vspace{5cm}}
\else
\centerline{
{\hspace{-1.5cm}}
\psfig{figure=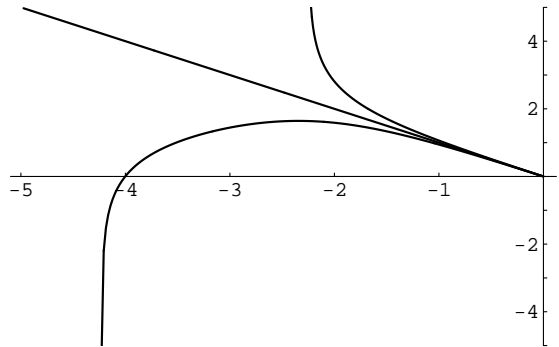,height=10.1cm,silent=}
\vspace{-2.2cm}}
\fi
{\caption{\baselineskip 3 ex
$\phi$ versus $z$ in units of $\chi$ with $\alpha = 1$.
The straight line in the middle
corresponds to the extreme solution. In the non-extreme
case with
$\beta=0.01$, the dilaton grows
without bound  shortly after $z = -2$.
For $\beta=-0.01$, corresponding to
the ultra-extreme case, the dilaton has a turning point and then
decreases without bound as the singularity
is approached.
\label{non_phi}}}
\end{figure}

\begin{figure}[htbp]
{\vspace{-1.7cm}}
\if\FiguresAvailable\NO
    {\vspace{5cm}}
\else
\centerline{
{\hspace{-1.5cm}}
\psfig{figure=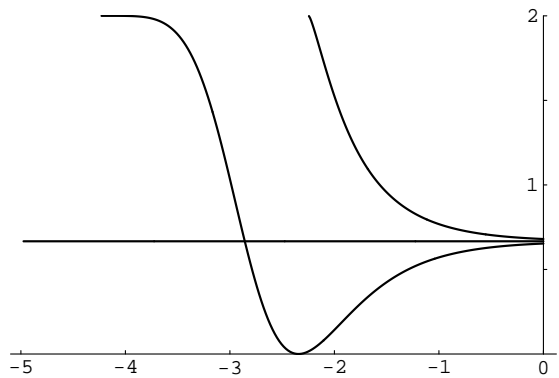,height=10.1cm,silent=}
\vspace{-2.2cm}}
\fi
{\caption{\baselineskip 3 ex
$\gamma$ versus $z$ in units of $\chi$ with $\alpha = 1$.
The straight line
corresponds to the extreme solution. In the non-extreme
case with
$\beta=0.01$, the $\gamma$ grows monotonically towards
the limit $\gamma=2$ as the singularity is approached.
The equation of state for $\beta=-0.01$ drops
down to $\gamma=0$ at the turning point of the scalar field,
and then
asymptotes towards $\gamma=2$ near the singularity.
\label{non_gamma}}}
\end{figure}

In order to  understand  these surprising  results,
it is instructive to
look at the evolution of the dilaton. The general behaviour is
illustrated in Fig.~\ref{non_phi}.
The extreme solutions with $0<\alpha<3$ are characterized by a delicate
balancing of the kinetic and potential energies.
As soon as the supersymmetry is broken, {\em i.e\/}.\
$\beta\ne 0$, the dilaton
speeds away up (non-extreme case)\
or down (ultra-extreme case)\ its potential.
As seen in Fig.~\ref{non_gamma},
the kinetic energy eventually  becomes dominant in both cases and is thus
responsible for the appearance of a naked singularity.

\paragraph*{The string frame and its generalizations:}
We would  also like to comment on the
physical significance of choosing a different
frame for description of the space--time in the domain wall background.
For the extreme domain walls there is a  choice of a conformal frame
defined by
\begin{equation}
\breve{g}_{\mu\nu}\equiv e^{2\phi/\sqrt{\alpha}}g_{\mu\nu}
\end{equation}
in which the space--time metric is {\em flat\/}\cite{Cvetic2}, {\em i.e\/}.\
$\breve{g}_{\mu\nu}=\eta_{\mu\nu}$.
For $\alpha=1$ this frame is known as the ``string frame'', {\em i.e\/}.\
the
frame in which
all the modes of the string theory couple coherently to the
dilaton field.  Since the  metric is flat in the string frame,
strings, which include all the   modes,   are ``blind''
to the curvature and to the singularities in the extreme
domain wall backgrounds.
One could then argue that the singularities are
artefacts coming from the use of the Einstein frame,
and that the fundamental physics as described
using the
metric $\breve{g}_{\mu\nu}=\eta_{\mu\nu}$, {\em i.e\/}.\
the string frame metric
and its generalizations,
is well behaved.

\begin{figure}[htbp]
{\vspace{-1.7cm}}
\if\FiguresAvailable\NO
    {\vspace{5cm}}
\else
\centerline{
{\hspace{-1.5cm}}
\psfig{figure=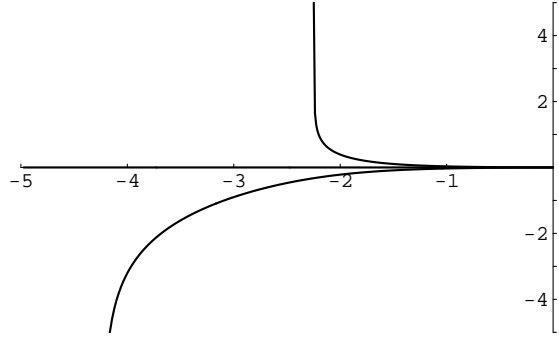,height=10.1cm,silent=}
\vspace{-2.2cm}}
\fi
{\caption{\baselineskip 3 ex
$\breve{a}$ versus $z$ in units of $\chi$ with $\alpha = 1$.
When $\beta=0$, $\breve{a}=0$ and the metric is flat.
In the ultra-extreme case $\beta=-0.01$, $\breve{a}\rightarrow -\infty$
and the conformal factor
goes to zero at finite $z$. In the non-extreme case with
$\beta=0.01$,
$\breve{a}\rightarrow\infty$, and
the conformal factor grows without bound at a finite value of
$z$.
\label{String_Frame}}}
\end{figure}

In the extreme case  there is always a frame   with
$\breve{g}_{\mu\nu}=\eta_{\mu\nu}$, and in which
(naked)\ singularities  are swept under the rug.
This, however,
is not possible in the non- and ultra-extreme cases.
In the non-extreme
case the conformal factor in the $\breve{g}_{\mu\nu}$-frame  grows as
one goes away from the wall, and in the ultra-extreme case
it decreases, see Fig.~\ref{String_Frame}.
This implies that  singularities are now felt in the
$\breve{g}_{\mu\nu}$-frame as well.

Furthermore, this result  seems to imply a paradox. According to
measurements in the Einstein frame,
the area of an embedding surface
at constant  co-moving
time decreases as one goes away from the wall
on the side with
the lowest value of the matter field potential. This is true
in both the non- and ultra-extreme
cases.
This side is therefore {\em inside\/} the bubble.
But now, according to the same type of measurements in the
$\breve{g}_{\mu\nu}$-frame, the
side with the lowest value of the matter field potential
is on the {\em outside\/} of the bubble
in the ultra-extreme case.
How can the same side be both on the inside and on the outside
of the wall? The
paradox is resolved when one realizes
that this local definition  of inside and outside
depends on the conformal frame and that, because of the singularity
cutting through space--time, there is no topological
obstruction to turn
it ``inside out.''

\subsection{Reflection-symmetric walls}
\label{Sub:Reflection}

Now we consider  a reflection-symmetric
non-extreme wall (a special case of Type~II
walls)\
with non-vanishing $V_{0}$ and a
running dilaton.\footnote{Of course, there
are no reflection-symmetric ultra-extreme walls.}
The potential is taken to be that of
Eq.~(\ref{potential})\ with $f(\phi)=e^{2\sqrt{\alpha}\phi}$
and $\widehat V(\phi)=0$.
According to Section~\ref{Sub:FieldAnsatze}, the boundary conditions
are
\begin{equation}
\left. \begin{array}{l}
\left.\phi'\right|
{\mbox{\raisebox{-1.2ex}{{\tiny{$\!0^{-}$}}}}}
=
-\left.\phi'\right|
{\mbox{\raisebox{-1.2ex}{{\tiny{$\!0^{+}$}}}}}
 = -\frac{1}{4}\sqrt{\alpha}\sigma,\\
\left.a'\right|
{\mbox{\raisebox{-1.2ex}{{\tiny{$\!0^{-}$}}}}}
=
-\left. a'\right|
{\mbox{\raisebox{-1.2ex}{{\tiny{$\!0^{+}$}}}}}
=
 \frac{1}{4}\sigma .
       \end{array}
\right.
\label{nonreflectionboundaryconditions}
\end{equation}
With these boundary conditions,
Eq.~(\ref{fieldeqs3})\ gives
\begin{equation}
-3 \beta^2
+
V_{0} + (3-\alpha) \left(\frac{\sigma}{4}\right)^2 =0.
\end{equation}
If $\alpha\neq 3$, we find by use of
Eq.~(\ref{chidetermined})\ that
\begin{equation}
\sigma = 4\sqrt{ \chi_1^2 + 3\, (3-\alpha )^{-1}\beta^2}.
\end{equation}
For $\alpha=0$, this expression reduces to the one found for
reflection-symmetric vacuum domain walls \cite{CGS}.
If $\alpha=3$, then $\sigma$ remains undetermined from this
expression, but
\begin{equation}
V_{0}= 3\beta^2.
\end{equation}
This result again indicates that in  the non-supersymmetric case  with
$\alpha=3$, the potential
itself is modified to be non-zero.
Thus, these walls are different from reflection-symmetric
domain walls in the background of a
Zel'dovich fluid \cite{Vuille}  or, which is equivalent,
domain walls minimally
coupled to a scalar field with no effective potential \cite{Wang}.
Yet, in all these cases one encounters
naked singularities.

\begin{figure}[htbp]
{\vspace{-1.7cm}}
\if\FiguresAvailable\NO
    {\vspace{5cm}}
\else
\centerline{
{\hspace{-1.5cm}}
\psfig{figure=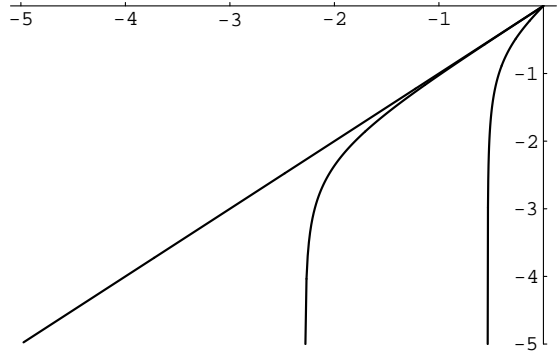,height=10.1cm,silent=}
\vspace{-2.2cm}}
\fi
{\caption{\baselineskip 3 ex
$a$ versus $z$ in units of $\chi$ for a reflection-symmetric
wall with $\alpha = 1$.
The straight line
corresponds to the extreme solution. The other two curves
represent non-extreme walls with
$\beta=0.1$ and $\beta=1$.
\label{non_ref_a}}}
\end{figure}

The boundary
conditions for reflection-symmetric non-extreme walls
are
different from those
of the wall adjacent to Minkowski space. In this case the
singularity is further away from the wall (compare
Figs.~\ref{non_a} and \ref{non_ref_a}); however,   as for the Type~I
non-extreme walls, the
reflection-symmetric solutions always have naked singularities as well.

\subsection{Non- and ultra-extreme domain walls
with a self-interacting dilaton}
\label{Sub:Self-interaction}

Let us now consider the situation where,
supersymmetry breaking
due to non-perturbative effects,
introduces an additional self-interaction term
in the dilaton potential. In general such a term is
of a complicated form.
One might hope that  a dilaton self-interaction term, providing
a mass for the dilaton, could stabilize the system
and keep the dilaton from running away.
For the sake of simplicity we
consider  an additional self-interaction potential $\widehat V$ of
the form
\begin{equation}
\widehat V = \lambda^2 \chi_1^2 \sinh^2 \omega \phi,
\end{equation}
where $\lambda$ and $\omega$ are real constants.
Note that in the cosmological
picture, this  potential has the opposite sign.

Since $V(0)=V'(0)=0$, the boundary  conditions for the
equations of motion at $z = 0^{-}$ for a wall
adjacent to Minkowski space, remain as
in Eq.~(\ref{boundarydynamical}).

\begin{figure}[htbp]
{\vspace{-1.7cm}}
\if\FiguresAvailable\NO
    {\vspace{5cm}}
\else
\centerline{
{\hspace{-1.5cm}}
\psfig{figure=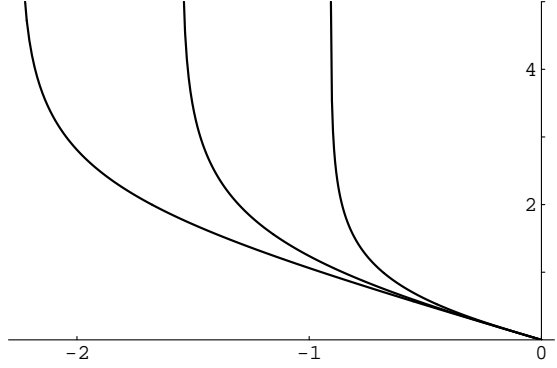,height=10.1cm,silent=}
\vspace{-2.2cm}}
\fi
{\caption{\baselineskip 3 ex
$\phi$ versus $z$ in units of $\chi$ with $\alpha = 1$
in  non-extreme solutions with $\beta=0.01$.
The left-most curve represents the
non-extreme solution with no self-interaction term ($\lambda=0$).
The other
two correspond to
$\lambda=1$ and $\omega=1$, and
$\lambda=1$ and $\omega=2$, respectively.
\label{non_phi_self}}}
\end{figure}

The physics of this problem is most easily understood in
the cosmological picture. We start by sending the dilaton up
the exponential potential. In the non-extreme case it
continues to roll up the potential until a singularity is reached.
Adding  a mass term, which would be negative in the cosmological picture,
makes the dilaton decelerate less and as a result it rolls faster
in the same direction. Hence, it is
clear that
such a self-interacting potential only
contributes to an earlier appearance of
the naked singularity in the non-extreme case.
Examples are depicted in Fig.~\ref{non_phi_self}.

\begin{figure}[htbp]
{\vspace{-1.7cm}}
\if\FiguresAvailable\NO
    {\vspace{5cm}}
\else
\centerline{
{\hspace{-1.5cm}}
\psfig{figure=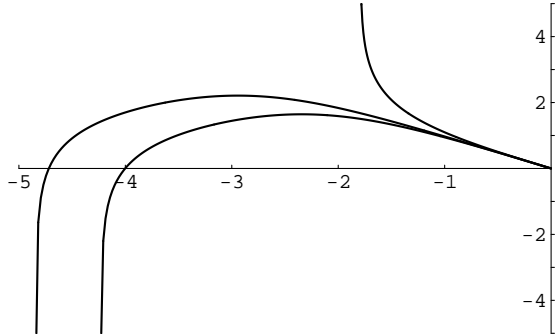,height=10.1cm,silent=}
\vspace{-2.2cm}}
\fi
{\caption{\baselineskip 3 ex
$\phi$ versus $z$ in units of $\chi$ with $\alpha = 1$
in ultra-extreme solutions with $\beta=-0.01$.
If the self-interaction is strong ($\lambda=1$ and $\omega=1$),
the dilaton
goes to positive infinity at the singularity. For
a weaker self-interaction ($\lambda=1$ and $\omega=0.45$),
the singularity comes later than in the case without
self-interaction (lower-most curve).
\label{ultra_phi_self}}}
\end{figure}

Now, in the ultra-extreme case,
the dilaton reaches a maximum in its exponential potential
and then returns. A self-interaction potential
would tend to accelerate the dilaton in the positive direction.
If this force is strong enough,
the result will be as in the non-extreme
case. If the exponential potential dominates, the dilaton will
again return. In either case, the dilaton runs off and produces
a naked singularity. In the ultra-extreme case,
this can only be avoided by a fine-tuned self-interaction
potential whose fine-tuning would have to depend also on $\beta$.
Examples of these two scenarios are shown in
Fig.~\ref{ultra_phi_self}.

The same qualitative features take place
in the reflection-symmetric case as well.

\subsection{Self-dual dilaton coupling}
\label{Sub:Self-Dual}

Let us now
assume that the  dilaton coupling $f(\phi)$ is self-dual. Namely,  for a finite
$\phi=\phi'$, $f(\phi)$ satisfies:
\begin{equation}
\left.{{\partial f}\over{\partial \phi}}\right|
{\mbox{\raisebox{-1.7ex}{{\tiny{$\!\phi'$}}}}}
=0
\label{fAnsatz}
\end{equation}
and that $\widehat V(\phi)=0$ (see Eq.~(\ref{dilpot})).

Note that the equation for  the dilaton
is of the form (see Eq.~(\ref{fieldeqs2}))
\begin{equation}
2{\phi}'' + 4 a '\phi'
+  e^{2a}\,{{\partial f(\phi)}\over{\partial\phi}} V_{0}=0,
\label{dilsd}
\end{equation}
with $\phi'=d\phi/dz$.  With  the boundary conditions
$\left.\phi\right|
{\mbox{\raisebox{-1.2ex}{{\tiny{$\!0$}}}}}=\phi'$ and
$\left.\phi'\right|
{\mbox{\raisebox{-1.2ex}{{\tiny{$\!0$}}}}}
= 0$,
the solution of the field equations corresponds
to a dilaton frozen at $\phi=\phi'$.
The global space--times
of these solutions are then identical to those
of the vacuum domain walls
\cite{CGS,CDGS}.

We would
like to address a stability of the constant dilaton
$\phi(z)=\phi'$ solutions for extreme, non- and ultra-extreme solutions.
In order to check the stability of such solutions,
one adds to
$\phi(0)=\phi'$
an
infinitesimal virtual displacement $\delta(0)$.
Consequently,
$\delta(z)$ satisfies the following equation:
\begin{equation}
2\delta'' +
4 a'\delta'+
e^{2a}\left.{{\partial^2\!f}\over{\partial \phi}^2}
\right|
{\mbox{\raisebox{-1.6ex}{{\tiny{$\!\phi'$}}}}}
V_{0}\delta  = 0
\label{expansion}
\end{equation}
 which was obtained  by expanding
$\phi(z)=\phi'+\delta(z)$ in Eq.~(\ref{dilsd})\ and
using
Eq.~(\ref{fAnsatz}).

In Subsection~\ref{Sub:Self-dual}
we have already shown that a supersymmetric
embedding of an extreme solution  with a constant
(complex)\ dilaton  $S=S'$
{\em always\/} renders it unstable.
We can also  show an
instability of such solutions  by
solving
Eq.~(\ref{expansion})\
directly,
 {\em i.e\/}.\
without reference to  the effective  dilaton
potential restricted by $N=1$ supergravity
theory.\footnote{The extreme solutions with a real dilaton field $\phi$
can be viewed as corresponding to
a special supersymmetric embedding, which renders
the imaginary part of the complex field $S$ constant.}
On the $z<0$ side of the wall  the
constant dilaton ($\phi=\phi'$)\
solution corresponds to
the  anti-de~Sitter  space--time. Thus, $a(z)$  is a solution of
Eq.~(\ref{fieldeqs1})\
with $\beta=0$ and
$\phi=\phi'$. It  is of the form:
\begin{equation}
a=-\ln\left(1-\chi_{1}z\right),
\end{equation}
where
$\chi_1^2=-f(\phi')V_0(\tau_1)/3$.
Eq.~(\ref{dilsd}) is then of the form:
\begin{equation}
\frac{d^2 \delta}{d y^2} -
3\frac{d\delta}{dy}-\frac{9}{4}b_{0}\delta=0,
\label{perturbationEq}
\end{equation}
where $b_0\equiv 2/3 (\partial^2 f/\partial\phi^2)f^{-1}|_{\phi'}$ and
$y=\ln(1-\chi_1z)$. The general form of the solution is
\begin{equation}
\delta(z)  ={\cal C}_1 (1-\chi_1z)^{\lambda_1}+{\cal C}_2
(1-\chi_1z)^{\lambda_2}
\end{equation}
with $\lambda_{1,2}=3\,( 1\pm\sqrt{1+b_0})/2$;
${\cal C}_{1,2}$ are
constants determined by initial conditions for $\delta(0)$.
Consequently, as
$z\rightarrow-\infty$,
$|\delta|\rightarrow \infty$ in both cases,
when the self-dual point  corresponds to the
maximum  ($b_0>0$)\ as well as  the
minimum of ($b_0<0$)\ the effective  dilaton potential.
Thus, the extreme self-dual
solutions with a frozen
dilaton are always dynamically unstable.

We would now like to
turn to the stability of the  non-extreme solutions
($\beta\ne 0$). On  the $z<0$ side
the space--time is
anti-de~Sitter with $a(z)$
(a solution of Eq.~(\ref{fieldeqs1})\ with
$\phi=\phi'$ and $\beta\ne 0$)\ of the form\cite{CGS}:
\begin{equation}
e^{2a(z)}=\left({\beta\over\chi_1}\right)\sinh^{-2}[\beta(z-z')]
\end{equation}
with $e^{2\beta z'}\equiv
1+2(\beta/\chi_1)^{2}+\beta/\chi_1\sqrt{1+(\beta/\chi_1)^2}$
and $\chi_1^2=-f(\phi')V_0(\tau_1)/3$.
Eq.~(\ref{expansion})\
for $\delta(z)$   can then be cast in the form:
\begin{equation}
\delta''-2\beta\coth[\beta(z-z')]\delta'
-
b_0\beta^2\sinh^{-2}[\beta(z-z')]\delta=0,
\label{nonexpansion}\end{equation}
where now $b_0\equiv (\partial^2 f/\partial\phi^2)f^{-1}|_{\phi'}$.
The general
solution for $\delta$  is complicated; however,
as $z\rightarrow-\infty$,  Eq.~(\ref{nonexpansion})\
reduces to:
\begin{equation}
w\frac{d^2 \delta}{d w^2}
+3\frac{d\delta}{dw}-4 b_{0}w \delta=0,
\label{perturbationEqp}
\end{equation}
where $w= e^{\beta z}$.
Hence, as $z\rightarrow -\infty$,
$\delta$ approaches a constant value, and thus the
solution {\em is
stable\/} under this perturbation.

Note that the same result applies to
the non-extreme solutions where the
constant dilaton solution
corresponds to the Minkowski space--time, {\em i.e\/}.\
$f(\phi')V_0(\tau_1)=0$. In this case
\begin{equation}
a(z)=\beta z
\end{equation}
and
Eq.~(\ref{expansion})\ is again
of the form
(\ref{perturbationEqp})\ with $w= e^{\beta
z}$ but  now $b_0\equiv -1/4\beta^{-2} (\partial^2
f/\partial\phi^2)|_{\phi'}V_0(\tau_1)$.
Clearly, as $z\rightarrow -\infty$,
$\delta$ approaches a constant value,
and therefore the solution is again
stable.

Along similar lines one can
prove that constant dilaton ultra-extreme solutions
are also stable against an infinitesimal  perturbation.
In  particular, in
the  case of Minkowski
constant dilaton vacuum on the $z>0$ side of the wall,
$a(z)=\beta z$ and
Eq.~(\ref{expansion})\
is of the form (\ref{perturbationEqp})\
with $w={\rm e}^{\beta
z}$ and   $b_0\equiv -1/4\beta^{-2} (\partial^2
f/\partial\phi^2)|_{\phi'}V_0(\tau_2)$.
The solution is of the form:
\begin{equation}
\delta(z)={\cal D}Y^{-1}{{{\cal J}_1(Y)}} ,
\label{ultrastab}
\end{equation}
where  ${\cal J}_1(Y)$ is the Bessel function of integer order one,
$Y=\sqrt{-4b_0} e^{\beta z}$ and ${\cal D}$ is a constant. As
$z\rightarrow \infty$, $Y\rightarrow \infty$ and  $\delta(z)\propto
Y^{-3/2}\cos(Y-3/4\pi)\rightarrow 0$. Thus, the ultra-extreme solution is
stable as well.

In conclusion,  in a theory with a self-dual dilaton coupling
extreme solutions  with a constant dilaton
are always unstable against small perturbation.
The origin of  the instability of
extreme  solutions
may be related to an infinite extent of such planar
configurations.  On the other hand,
non- as well as
ultra-extreme solutions
with a constant dilaton are  always
dynamically stable.

\section{Conclusion}
\label{Sect:Conclusion}

We have analysed the gravitational fields induced by
dilatonic domain walls with an exponential dilaton coupling
and with a self-dual dilaton coupling.

We have shown that generic non- and ultra-extreme dilatonic domain walls
with an exponential dilaton coupling ($e^{2\sqrt\alpha\phi}$)\ have
naked singularities. In contrast, extreme walls  have planar
naked singularities for $\alpha>1$, while for $\alpha\le 1$ the
singularities are null.
In the  frame where the extreme domain wall metric is flat
(and which for
$\alpha=1$ corresponds to the string frame),
the non-extreme domain walls still have naked singularities,
whereas the
ultra-extreme singularities
are hidden behind horizons in this frame.

The ultra-extreme domain walls correspond to false vacuum decay
bubbles and as such they might have a direct
physical relevance if the Minkowski  vacuum turns out to be
unstable.
A vacuum decay would, according to the theory we have used,
force the dilaton to start running even if it
had
previously
been
trapped by a mass term or another self-interaction potential.
The kinetic
energy of the
dilaton would then grow without bound and lead to a naked
singularity.

For  exponential  dilaton coupling the non- and ultra-extreme
domain walls have naked singularities, which
cannot be avoided by adding a mass term for the dilaton.
On the other hand, when the dilaton coupling is self-dual,
there are singularity-free domain wall configurations:
outside
the domain wall region, the dilaton is
then trapped at the self-dual point and
the domain wall geometries reduce to those of
the singularity-free
vacuum domain walls. This class of solutions
is dynamically stable  for non- and ultra-extreme walls, but
the extreme solutions of such a type are not dynamically stable.

We have therefore arrived at the conclusion that for a
large class of theories with
dilaton(s), the space--times of  non- and ultra-extreme  domain walls
(as well as some  extreme domain walls)\
are necessarily plagued
with naked singularities.
If one believes that the gravitational field of vacuum decay bubbles
and non-extreme domain walls should be singularity-free, this
observation imposes serious constraints on the phenomenological
viability of such theories. In particular,
the exponential dilaton coupling
of the effective tree level  action from superstrings is ruled out.
The results obtained indicate that
non-perturbative effects,
{\em e.g\/}.\  non-perturbatively induced  dilaton
superpotential within supergravity theories
and/or a  non-perturbatively induced
 self-dual   dilaton K\" ahler potential,
should  play a  crucial r\^ole
in altering qualitatively  the
space--time structure  of the dilatonic domain
wall space--times.

\acknowledgments{
This work is supported by the U.S. DOE
Grant No. DOE-EY-76-02-3071 (M.C.) and a NATO Collaborative Research Grant
(M.C.). M.C. would like
to thank D. Youm for useful discussions and CERN, where part of this work was
done,
for hospitality.
H.H.S. thanks the University of Pennsylvania, where this
work was initiated, for
its
hospitality.
We acknowledge the use of {\em Mathematica\/}
and the tensor package {\sc Cartan}. }



\end{document}